\documentclass[a4paper,12pt]{article}
\usepackage{amssymb}
\usepackage{amsmath}
\usepackage{epsfig}
\usepackage{subfigure}
\usepackage{graphics}
\usepackage{tikz}
\usepackage{latexsym}
\usepackage{rotating}
\usepackage{titlesec}
\usetikzlibrary{matrix}

\title{Hirota bilinear forms of the AKNS($N$) systems}

\author{Metin G\"{u}rses \thanks{gurses@fen.bilkent.edu.tr}\\
{\small Department of Mathematics, Faculty of Science}\\
{\small Bilkent University, 06800 Ankara - Turkey}\\
Asl{\i} Pekcan \thanks{aslipekcan@hacettepe.edu.tr} \\
{\small Department of Mathematics, Faculty of Science} \\
{\small Hacettepe University, 06800 Ankara - Turkey}
}

\date{\nonumber}
\begin{document}
\maketitle
\date{\nonumber}
\begin{abstract}
We study the AKNS($N$) hierarchy for $N=3,4,5,6$. We give the Hirota bilinear forms of these systems and present local and nonlocal reductions of them. We give the Hirota bilinear forms of the reduced equations. The compatibility of the commutativity diagrams of the application of the recursion operator, reductions of the AKNS($N$) systems, and Hirota bilinearization is also studied.

\end{abstract}

\section{Introduction}

There are several works about finding new integrable nonlocal nonlinear partial differential equations and obtaining different kinds of solutions of these nonlocal equations after Ablowitz and Musslimani introduced nonlocal type of reductions \cite{AbMu1}-\cite{AbMu3}. If these nonlocal reductions are done consistently, the integrable systems can be reduced to the reverse space, the reverse time, and the reverse space-time nonlocal equations, which are also integrable. Particularly, there are works on nonlocal nonlinear Schr\"{o}dinger equations (NLS) \cite{AbMu1}-\cite{jianke},
nonlocal modified Korteweg-de Vries (mKdV) equations \cite{AbMu2}-\cite{chen}, \cite{GurPek3}, \cite{GurPek2}-\cite{ma},  nonlocal sine-Gordon (SG) equations \cite{AbMu2}-\cite{chen}, \cite{aflm}, and so on \cite{fok}-\cite{hydro}.\\

The systems admitting nonlocal reductions have discrete symmetry transformations leaving the systems invariant. In \cite{origin} we showed that a special case of discrete symmetry transformations are actually the nonlocal reductions of the same systems. The connection between local and nonlocal reductions was given in \cite{Vincent}, \cite{Yang}.\\

 The AKNS hierarchy \cite{AKNS} can be written as
\begin{equation}\label{generalsys}
au_{t_N}=\mathcal{R}^{N} u_x \,\, \mathrm{where} \,\,  u= \left( \begin{array}{c}
p  \\
q
 \end{array} \right) \,\, \mathrm{i.e.} \,\,  \left( \begin{array}{c}
p_{t_N}  \\
q_{t_N}

 \end{array} \right)= \mathcal{R}^{N} \left( \begin{array}{c}
p_x  \\
q_x
 \end{array} \right),
\end{equation}
\noindent for $N=1,2,\cdots$, where $\mathcal{R}$ is the recursion operator,
\begin{equation}\label{recursion}
 \mathcal{R}=\left( \begin{array}{cc}
2pD^{-1}q-D & 2pD^{-1}p  \\
-2qD^{-1}q & -2qD^{-1}p+D
 \end{array} \right),
\end{equation}
\noindent and $a$ is an arbitrary constant. Here $D$ is the total $x$-derivative and $D^{-1}=\int^x$ is the standard anti-derivative.\\

 For $N=1$ the system (\ref{generalsys}) gives the coupled NLS system,
which we call the AKNS($1$) system,
\begin{align}
&ap_{t_1}=-p_{xx}+2p^2q,\label{n=1p}\\
&aq_{t_1}=q_{xx}-2pq^2.\label{n=1q}
\end{align}
Letting $p=\frac{g}{f}$ and $q=\frac{h}{f}$ in (\ref{n=1p}) and (\ref{n=1q}) yields the Hirota bilinear form of this system as
\begin{align}
&(aD_{t_1}+D_x^2)\{g\cdot f\}=0,\label{NLShirotaa}\\
&(aD_{t_1}-D_x^2)\{h\cdot f\}=0,\label{NLShirotab}\\
&D_x^2\{f\cdot f\}=-2gh.\label{NLShirotac}
\end{align}
We have studied soliton solutions, local and nonlocal reductions of this system in \cite{GurPek1}, \cite{GurPek3}. Similarly if we consider
the case for $N=2$ we have the coupled mKdV system, which we call the AKNS($2$) system,
\begin{align}
&ap_{t_2}=-p_{xxx}+6pqp_x,\label{n=2p}\\
&aq_{t_2}=-q_{xxx}+6pqq_x.\label{n=2q}
\end{align}
If we let $p=\frac{g}{f}$ and $q=\frac{h}{f}$ in (\ref{n=2p}) and (\ref{n=2q}), we get
the Hirota bilinear form of the system (\ref{n=2p}) and (\ref{n=2q})
as
\begin{align}
&(aD_{t_2}+D_x^3)\{g\cdot f\}=0,\label{mKdVhirotaa}\\
&(aD_{t_2}+D_x^3)\{h\cdot f\}=0,\label{mKdVhirotab}\\
&D_x^2\{f\cdot f\}=-2gh.\label{mKdVhirotac}
\end{align}
We have studied soliton solutions, local and nonlocal reductions of this system in \cite{GurPek3}, \cite{GurPek2}.\\

 For general $N+1$, the Hirota bilinear form of AKNS($N+1$) system can be represented by \cite{Newell}
\begin{align}
&(aD_{t_{N+1}}-D_xD_{t_N})\{g\cdot f\}=0,\label{AKNS(N+1)-a}\\
&(aD_{t_{N+1}}-D_xD_{t_N})\{h\cdot f\}=0,\label{AKNS(N+1)-b}\\
&D_x^2\{f\cdot f\}=-2gh.\label{AKNS(N+1)-c}
\end{align}
This is a recurrence relation for the Hirota bilinear forms of the AKNS hierarchy. Double Wronskian solutions of the AKNS hierarchy expressed by (\ref{AKNS(N+1)-a})-(\ref{AKNS(N+1)-c}) are obtained in \cite{QML} and \cite{MPQY}. Chen et al. \cite{chen} gave nonlocal reductions of the AKNS hierarchy (\ref{AKNS(N+1)-a})-(\ref{AKNS(N+1)-c}). They also derived exact solutions in double Wronskian form of the reduced nonlocal equations from those double Wronskian solutions of the AKNS hierarchy. Ankiewicz et al. \cite{Anki} analyzed the infinite hierarchy of AKNS equations in two parts; even-order members and odd-order members. They gave the generalized soliton solutions, plane wave solutions, Kuznetsov-Ma breathers, Akhmediev
breathers, rogue and periodic
wave solutions for this hierarchy.\\

In studying Hirota bilinear forms and reduction of AKNS($N$) systems we face up with the commutativity of certain operations, such as application of the recursion operator and then reduction of systems illustrated in the diagram below.

\begin{figure}[h]
\centering
\begin{tikzpicture}
  \matrix (m) [matrix of math nodes,row sep=3em,column sep=4em,minimum width=2em]
  {
     \mathrm{AKNS}(N) & \mathrm{AKNS}(N+1) \\
     \mathrm{AKNS_{red}}(N) & \mathrm{AKNS_{red}}(N+1) \\};
  \path[-stealth]
    (m-1-1) edge node [left] {Reduction} (m-2-1)
            edge [] node [below] {$\mathcal{R}$} (m-1-2)
    (m-2-1.east|-m-2-2) edge node [below] {$\mathcal{R}_{red}$}
            node [above] {} (m-2-2)
    (m-1-2) edge node [right] {Reduction} (m-2-2)
           ;
           \end{tikzpicture}
           \caption{Relation between reductions and hierarchy}
\end{figure}
\noindent Here $\mathcal{R}$ is the recursion operator (\ref{recursion}), $\mathcal{R}_{red}$ is the same operator with a reduction applied, and
$\mathrm{AKNS_{red}}(N)$ is the AKNS($N$) system (\ref{generalsys}) with a reduction applied. In general, we have the recursion operator, reduction of the system, and Hirota bilinearization. We study the compatibility of the commutativity diagrams for each pair of the operators.

In this paper we study the system AKNS($N$) given by (\ref{generalsys}) for higher orders, particularly for $N=3,4,5,6$. The recurrence relation (\ref{AKNS(N+1)-a})-(\ref{AKNS(N+1)-c}) is not very useful to obtain the Hirota bilinear form of the system AKNS($N$) for $N\geq 3$. For $N\geq 3$, the Hirota bilinear form for the system AKNS($N$) is not as simple as these for $N=1, 2$. We show that the Hirota bilinear forms of AKNS($N$) systems are not homogeneous for $N=3,4,5,6$. In Section 2, we give the Hirota bilinear forms of AKNS($3$) and AKNS($4$) systems. Then we consider the local and nonlocal reductions of these systems. We present the Hirota bilinear forms of the reduced equations. Similarly, in Section 3 we study the AKNS($5$) and AKNS($6$) systems. In Section 4, we analyze the commutativity of three diagrams giving the relations between the reductions, recursion operators, bilinearization, and soliton solutions.

\section{Hirota bilinear forms of AKNS($N$) systems for $N=3, 4$}

\noindent \textbf{A. AKNS($N$) system for $N=3$}\\

\noindent For $N=3$,  we have the AKNS($3$) system
\begin{align}
&ap_{t_3}=-p_{xxxx}+6qp_x^2+4pp_xq_x+8pqp_{xx}+2p^2q_{xx}-6p^3q^2,\label{n=3p}   \\
&aq_{t_3}=q_{xxxx}-6pq_x^2-4qp_xq_x-8pqq_{xx}-2q^2p_{xx}+6p^2q^3.\label{n=3q}
\end{align}
If we let $p=\frac{g}{f}$ and $q=\frac{h}{f}$ in (\ref{n=3p}) and (\ref{n=3q}), we get the Hirota bilinear form of AKNS($3$) system,
\begin{align}
&(aD_{t_3}+D_x^4)\{g\cdot f\}=-3hs,\label{n=3-1}\\
&(aD_{t_3}-D_x^4)\{h\cdot f\}=3g\tau,\label{n=3-2}\\
& D_x^2\{g\cdot g\}=fs,\label{n=3-3}\\
&D_x^2\{h\cdot h\}=f\tau,\label{n=3-4}\\
& D_x^2\{f\cdot f \}=-2gh ,\label{n=3-5}
\end{align}
where $s$ and $\tau$ are auxiliary functions.\\

\noindent \textbf{a. Local reductions for AKNS($3$) system}\\

\noindent The AKNS($3$) system does not have $q(x,t_3)=k p(x,t_3)$ type local reduction but we can consider
the local reduction $q(x,t_3)=k\bar{p}(x,t_3)$, $k$ is a real constant, for this case.

\noindent \textbf{a.i.} $q(x,t_3)=k\bar{p}(x,t_3)$, $k$ is a real constant.\\

\noindent Under this reduction the AKNS($3$) system (\ref{n=3p}) and (\ref{n=3q})
consistently reduces to the local complex AKNS($3$) equation
\begin{equation}\label{localcomplexAKNS(3)}
ap_{t_3}+p_{xxxx}-6k\bar{p}p_x^2-4kpp_x\bar{p}_x-8kp\bar{p}p_{xx}-2kp^2\bar{p}_{xx}+6k^2\bar{p}^2p^3=0,
\end{equation}
where $\bar{a}=-a$. For $k=-1$ and $a=i$, this equation is known as fourth order NLS equation or Lakshmanan-Porsezian-Daniel equation \cite{YSC}-\cite{LPD3}.\\

\noindent We can also obtain the Hirota bilinear form of the reduced equation (\ref{localcomplexAKNS(3)}).
From the reduction $q(x,t_3)=k\bar{p}(x,t_3)$ we have
\begin{equation}
\frac{h(x,t_3)}{f(x,t_3)}=k\frac{\bar{g}(x,t_3)}{\bar{f}(x,t_3)}.
\end{equation}
By equating numerator and denominator separately that is by using Type 1 approach \cite{GurPek1}, \cite{GurPek3}, \cite{GurPek2} we have
$h(x,t_3)=k\bar{g}(x,t_3)$ and $f(x,t_3)=\bar{f}(x,t_3)$. When we use these relations in (\ref{n=3-1})-(\ref{n=3-5}) we get the Hirota bilinear form of
the reduced local complex AKNS($3$) equation (\ref{localcomplexAKNS(3)}) as
\begin{align}
&(aD_{t_3}+D_x^4)\{g\cdot f\}=-3k\bar{g}s,\label{locn=3-1}\\
& D_x^2\{g\cdot g\}=fs,\label{locn=3-2}\\
& D_x^2\{f\cdot f \}=-2k|g|^2,\label{locn=3-3}
\end{align}
where $a$ is a pure imaginary number.\\

\noindent \textbf{b. Nonlocal reductions for AKNS($3$) system}\\

\noindent \textbf{b.i.} $q(x,t_3)=kp(\varepsilon_1x,\varepsilon_2t_3)=kp^{\varepsilon}$, $\varepsilon_1^2=\varepsilon_2^2=1$, $k$ is a real constant.\\

\noindent When we apply this reduction to the AKNS($3$) system (\ref{n=3p}) and (\ref{n=3q}), it occurs that to have a consistent reduction of this type we must have $a\varepsilon_2=-a$
that is $\varepsilon_2=-1$. Therefore  we have the following nonlocal reduced equations:
\begin{align}\label{nonlocalrealAKNS(3)general}
&ap_{t_3}(x,t_3)=-p_{xxxx}(x,t_3)+6kp(\varepsilon_1x,-t_3)p_x^2(x,t_3)+4kp(x,t_3)p_x(x,t_3)p_x(\varepsilon_1x,-t_3)
\nonumber\\&+8kp(x,t_3)p(\varepsilon_1x,-t_3)p_{xx}(x,t_3)
+2kp^2(x,t_3)p_{xx}(\varepsilon_1x,-t_3)-6k^2p^3(x,t_3)p^2(\varepsilon_1x,-t_3),
\end{align}
which are written explicitly as\\

\noindent \textbf{(1)}\, $(\varepsilon_1,\varepsilon_2)=(1,-1)$ Reverse time nonlocal  AKNS($3$) equation:
\begin{align}\label{nonlocalrealAKNS(3)T}
&ap_{t_3}(x,t_3)=-p_{xxxx}(x,t_3)+6kp(x,-t_3)p_x^2(x,t_3)+4kp(x,t_3)p_x(x,t_3)p_x(x,-t_3)
\nonumber\\&+8kp(x,t_3)p(x,-t_3)p_{xx}(x,t_3)
+2kp^2(x,t_3)p_{xx}(x,-t_3)-6k^2p^3(x,t_3)p^2(x,-t_3).
\end{align}
\vspace{0.2cm}
\noindent \textbf{(2)}\, $(\varepsilon_1,\varepsilon_2)=(-1,-1)$ Reverse space-time nonlocal AKNS($3$) equation:
\begin{align}\label{nonlocalrealAKNS(3)ST}
&ap_{t_3}(x,t_3)=-p_{xxxx}(x,t_3)+6kp(-x,-t_3)p_x^2(x,t_3)+4kp(x,t_3)p_x(x,t_3)p_x(-x,-t_3)
\nonumber\\&+8kp(x,t_3)p(-x,-t_3)p_{xx}(x,t_3)
+2kp^2(x,t_3)p_{xx}(-x,-t_3)-6k^2p^3(x,t_3)p^2(-x,-t_3).
\end{align}

\noindent From the reduction, Type 1 approach gives $h(x,t_3)=kg(\varepsilon_1x,t_3)=kg^{\varepsilon}$ and $f(x,t_3)=f(\varepsilon_1x,-t_3)$.
Using these relations in (\ref{n=3-1})-(\ref{n=3-5}) yields the Hirota bilinear forms of the equations given by (\ref{nonlocalrealAKNS(3)general}) as
\begin{align}
&(aD_{t_3}+D_x^4)\{g\cdot f\}=-3kg^{\varepsilon}s,\label{nonlocal1n=3-1}\\
&D_x^2\{g\cdot g\}=fs,\label{nonlocal1n=3-2}\\
&D_x^2\{f\cdot f\}=-2kgg^{\varepsilon},\label{nonlocal1n=3-3}
\end{align}
where $s$ is an auxiliary function and $\varepsilon_1=\pm 1, \varepsilon_2=-1$.\\
\bigskip

\noindent \textbf{b.ii.} $q(x,t_3)=k\bar{p}(\varepsilon_1x,\varepsilon_2t_3)=k\bar{p}^{\varepsilon}$, $\varepsilon_1^2=\varepsilon_2^2=1$, $k$ is a real constant.\\

\noindent When we use this reduction on the AKNS($3$) system (\ref{n=3p}) and (\ref{n=3q}), it yields that to have a consistent reduction of this type we must have $\bar{a}\varepsilon_2=-a$. We get the nonlocal reduced equations,
\begin{align}\label{nonlocalcomplexAKNS(3)general}
&ap_{t_3}(x,t_3)=-p_{xxxx}(x,t_3)+6k\bar{p}(\varepsilon_1x,\varepsilon_2t_3)p_x^2(x,t_3)+4kp(x,t_3)p_x(x,t_3)\bar{p}_x(\varepsilon_1x,\varepsilon_2t_3)
\nonumber\\&+8kp(x,t_3)\bar{p}(\varepsilon_1x,\varepsilon_2t_3)p_{xx}(x,t_3)
+2kp^2(x,t_3)\bar{p}_{xx}(\varepsilon_1x,\varepsilon_2t_3)-6k^2p^3(x,t_3)\bar{p}^2(\varepsilon_1x,\varepsilon_2t_3).
\end{align}

\noindent Explicitly, we have three nonlocal reduced equations:\\

\noindent \textbf{(1)}\, $(\varepsilon_1,\varepsilon_2)=(-1,1)$ Reverse space nonlocal complex AKNS($3$) equation:
\begin{align}\label{nonlocalcomplexAKNS(3)S}
&ap_{t_3}(x,t_3)=-p_{xxxx}(x,t_3)+6k\bar{p}(-x,t_3)p_x^2(x,t_3)+4kp(x,t_3)p_x(x,t_3)\bar{p}_x(-x,t_3)
\nonumber\\&+8kp(x,t_3)\bar{p}(-x,t_3)p_{xx}(x,t_3)
+2kp^2(x,t_3)\bar{p}_{xx}(-x,t_3)-6k^2p^3(x,t_3)\bar{p}^2(-x,t_3),
\end{align}
where $a$ is a pure imaginary number.\\

\noindent \textbf{(2)}\, $(\varepsilon_1,\varepsilon_2)=(1,-1)$ Reverse time nonlocal complex AKNS($3$) equation:
\begin{align}\label{nonlocalcomplexAKNS(3)T}
&ap_{t_3}(x,t_3)=-p_{xxxx}(x,t_3)+6k\bar{p}(x,-t_3)p_x^2(x,t_3)+4kp(x,t_3)p_x(x,t_3)\bar{p}_x(x,-t_3)
\nonumber\\&+8kp(x,t_3)\bar{p}(x,-t_3)p_{xx}(x,t_3)
+2kp^2(x,t_3)\bar{p}_{xx}(x,-t_3)-6k^2p^3(x,t_3)\bar{p}^2(x,-t_3),
\end{align}
where $a \in \mathbb{R}$.\\

\noindent \textbf{(3)}\, $(\varepsilon_1,\varepsilon_2)=(-1,-1)$ Reverse space-time nonlocal complex AKNS($3$) equation:
\begin{align}\label{nonlocalcomplexAKNS(3)ST}
&ap_{t_3}(x,t_3)=-p_{xxxx}(x,t_3)+6k\bar{p}(-x,-t_3)p_x^2(x,t_3)+4kp(x,t_3)p_x(x,t_3)\bar{p}_x(-x,-t_3)
\nonumber\\&+8kp(x,t_3)\bar{p}(-x,-t_3)p_{xx}(x,t_3)
+2kp^2(x,t_3)\bar{p}_{xx}(-x,-t_3)-6k^2p^3(x,t_3)\bar{p}^2(-x,-t_3),
\end{align}
where $a \in \mathbb{R}$.\\

\noindent From the reduction, Type 1 gives $h(x,t_3)=k\bar{g}(\varepsilon_1x,\varepsilon_2t_3)=k\bar{g}^{\varepsilon}$ and $f(x,t_3)=\bar{f}(\varepsilon_1x,\varepsilon_2 t_3)$. Hence the Hirota bilinear forms of the equations expressed by (\ref{nonlocalcomplexAKNS(3)general}) are given by
\begin{align}
&(aD_{t_3}+D_x^4)\{g\cdot f\}=-3k\bar{g}^{\varepsilon}s,\label{nonlocal2n=3-1}\\
&D_x^2\{g\cdot g\}=fs,\label{nonlocal2n=3-2}\\
&D_x^2\{f\cdot f\}=-2kg\bar{g}^{\varepsilon},\label{nonlocal2n=3-3}
\end{align}
where $s$ is an auxiliary function and $\bar{a}\varepsilon_2=-a$. \\

\noindent \textbf{B. AKNS($N$) system for $N=4$}\\

\noindent For $N=4$, we have the following system from (\ref{generalsys}) that we call AKNS($4$) system:
\begin{align}
&ap_{t_4}=-p_{xxxxx}+10pqp_{xxx}+10pp_xq_{xx}-30p^2q^2p_x+10pp_{xx}q_x+20qp_xp_{xx}+10p_x^2q_x,\label{n=4p}\\
&aq_{t_4}=-q_{xxxxx}+10pqq_{xxx}+10qq_xp_{xx}-30p^2q^2q_x+10qq_{xx}p_x+20pq_xq_{xx}+10p_xq_x^2.\label{n=4q}
\end{align}
Letting $p=\frac{g}{f}$ and $q=\frac{h}{f}$ in (\ref{n=4p}) and  (\ref{n=4q}) yields
\begin{align}
&(aD_{t_4}+D_x^5)\{g\cdot f\}=5D_x\{h\cdot s\},\label{n=4-1}\\
&(aD_{t_4}+D_x^5)\{h\cdot f\}=5D_x\{g\cdot \tau\},\label{n=4-2}\\
& D_x^2\{f\cdot f \}=-2gh, \label{n=4-3}\\
& D_x^2\{g\cdot g\}=fs,\label{n=4-4}\\
&D_x^2\{h\cdot h\}=f \tau, \label{n=4-5}
\end{align}
where $s$ and $\tau$ are auxiliary functions.\\

\noindent \textbf{a. Local reductions for AKNS($4$) system}\\

\noindent \textbf{a.i.} $q(x,t_4)=kp(x,t_4)$, $k$ is a real constant.\\

\noindent  Different than the AKNS($3$) system, the AKNS($4$) system (\ref{n=4p}) and (\ref{n=4q}) has such type of local
reduction consistently. The reduced local AKNS($4$) equation is
\begin{equation}\label{localrealAKNS(4)}
ap_{t_4}=-p_{xxxxx}+10kp^2p_{xxx}+10kpp_xp_{xx}-30k^2p^4p_x+10kpp_xp_{xx}+20kpp_xp_{xx}+10kp_x^3,
\end{equation}
without any constraint on the parameters. By Type 1, we obtain the Hirota bilinear form of this reduced local equation as
\begin{align}
&(aD_{t_4}+D_x^5)\{g\cdot f\}=5kD_x\{g\cdot s\},\label{loc1n=4-1}\\
& D_x^2\{g\cdot g\}=fs,\label{loc1n=4-2}\\
& D_x^2\{f\cdot f \}=-2kg^2, \label{loc1n=4-3}
\end{align}
where $s$ is an auxiliary function.\\

\noindent \textbf{a.ii.} $q(x,t_4)=k\bar{p}(x,t_4)$, $k$ is a real constant.\\

\noindent When we apply this reduction to the AKNS($4$) system (\ref{n=4p}) and (\ref{n=4q}) it
consistently reduces to the local complex AKNS($4$) equation
\begin{equation}\label{localcomplexAKNS(4)}
ap_{t_4}=-p_{xxxxx}+10kp\bar{p}p_{xxx}+10kpp_x\bar{p}_{xx}-30k^2p^2\bar{p}^2p_x+10kpp_{xx}\bar{p}_x+20k\bar{p}p_xp_{xx}+10kp_x^2\bar{p}_x,
\end{equation}
where $\bar{a}=a$. This equation for $a=1, k=-1$ is named as fifth-order NLS equation. Soliton, breather, and rogue wave solutions of this equation were considered in \cite{fifthNLS1}-\cite{fifthNLS7}. When we use Type 1 approach on (\ref{n=4-1})-(\ref{n=4-5}) with this reduction we obtain the Hirota bilinear form of the local complex AKNS($4$) equation (\ref{localcomplexAKNS(4)}) as,
\begin{align}
&(aD_{t_4}+D_x^5)\{g\cdot f\}=5kD_x\{\bar{g}\cdot s\},\label{loc2n=4-1}\\
& D_x^2\{g\cdot g\}=fs,\label{loc2n=4-2}\\
& D_x^2\{f\cdot f \}=-2k|g|^2, \label{loc2n=4-3}
\end{align}
where $s$ is an auxiliary function and $a\in \mathbb{R}$. Here $f(x,t_4)=\bar{f}(x,t_4)$.\\

\noindent \textbf{b. Nonlocal reductions for AKNS($4$) system}\\

\noindent \textbf{b.i.} $q(x,t_4)=kp(\varepsilon_1x,\varepsilon_2t_4)=kp^{\varepsilon}$, $\varepsilon_1^2=\varepsilon_2^2=1$, $k$ is a real constant.\\

If we use this reduction on the AKNS($4$) system (\ref{n=4p}) and (\ref{n=4q}) we get the condition $a\varepsilon_1\varepsilon_2=a$ i.e. $\varepsilon_1=\varepsilon_2=-1$ to have a consistent nonlocal reduction. Hence we obtain the following reverse space-time nonlocal  AKNS($4$) equation:
\begin{align}\label{nonlocalrealAKNS(4)ST}
&ap_{t_4}(x,t_4)=-p_{xxxxx}(x,t_4)+10kp(x,t_4)p(-x,-t_4)p_{xxx}+10kp(x,t_4)p_x(x,t_4)p_{xx}(-x,-t_4)
\nonumber\\
&-30k^2p^2(x,t_4)p^2(-x,-t_4)p_x+10kp(x,t_4)p_{xx}(x,t_4)p_x(-x,-t_4)\nonumber\\
&+20kp(-x,-t_4)p_x(x,t_4)p_{xx}(x,t_4)+10kp_x^2(x,t_4)p_x(-x,-t_4).
\end{align}

\noindent Using Type 1 gives the Hirota bilinear form of the above equation as
\begin{align}
& (aD_{t_4}+D_x^5)\{g\cdot f\}=5kD_x\{g^{\varepsilon}\cdot s\},\\
&D_x^2\{g\cdot g\}=fs,\\
&D_x^2\{f\cdot f\}=-2kgg^{\varepsilon},
\end{align}
where $s$ is an auxiliary function, $g^{\varepsilon}=g(-x,-t_4)$, and $f(x,t_4)=f(-x,-t_4)$.\\

\noindent \textbf{b.ii.} $q(x,t_4)=k\bar{p}(\varepsilon_1x,\varepsilon_2t_4)=k\bar{p}^{\varepsilon}$, $\varepsilon_1^2=\varepsilon_2^2=1$, $k$ is a real constant.\\

\noindent When we use this nonlocal reduction on the AKNS($4$) system (\ref{n=4p}) and (\ref{n=4q}), it occurs that to have a consistent reduction we must have $a=\bar{a}\varepsilon_1\varepsilon_2$. We get the nonlocal reduced equations,
\begin{align}\label{nonlocalcomplexAKNS(4)general}
&ap_{t_4}(x,t_4)=-p_{xxxxx}(x,t_4)+10kp\bar{p}(\varepsilon_1x,\varepsilon_2t_4)p_{xxx}(x,t_4)+10kp(x,t_4)p_x(x,t_4)\bar{p}_{xx}(\varepsilon_1x,\varepsilon_2t_4)
\nonumber\\&-30k^2p^2(x,t_4)\bar{p}^2(\varepsilon_1x,\varepsilon_2t_4)p_x(x,t)+10kp(x,t_4)p_{xx}(x,t_4)\bar{p}_x(\varepsilon_1x,\varepsilon_2t_4)
\nonumber\\&+20k\bar{p}(\varepsilon_1x,\varepsilon_2t_4)p_x(x,t_4)p_{xx}(x,t_4)+10kp_x^2(x,t_4)\bar{p}_x(\varepsilon_1x,\varepsilon_2t_4).
\end{align}
\noindent Explicitly, we have three nonlocal reduced equations.\\

\noindent \textbf{(1)}\, $(\varepsilon_1,\varepsilon_2)=(-1,1)$ Reverse space nonlocal complex AKNS($4$) equation:
\begin{align}\label{nonlocalcomplexAKNS(4)S}
&ap_{t_4}(x,t_4)=-p_{xxxxx}(x,t_4)+10kp\bar{p}(-x,t_4)p_{xxx}(x,t_4)+10kp(x,t_4)p_x(x,t_4)\bar{p}_{xx}(-x,t_4)
\nonumber\\&-30k^2p^2(x,t_4)\bar{p}^2(-x,t_4)p_x(x,t_4)+10kp(x,t_4)p_{xx}(x,t_4)\bar{p}_x(-x,t_4)
\nonumber\\&+20k\bar{p}(-x,t_4)p_x(x,t_4)p_{xx}(x,t_4)+10kp_x^2(x,t_4)\bar{p}_x(-x,t_4),
\end{align}
where $a$ is a pure imaginary number.\\

\noindent \textbf{(2)}\, $(\varepsilon_1,\varepsilon_2)=(1,-1)$ Reverse time nonlocal complex AKNS($4$) equation:
\begin{align}\label{nonlocalcomplexAKNS(4)T}
&ap_{t_4}(x,t_4)=-p_{xxxxx}(x,t_4)+10kp\bar{p}(x,-t_4)p_{xxx}(x,t_4)+10kp(x,t_4)p_x(x,t_4)\bar{p}_{xx}(x,-t_4)
\nonumber\\&-30k^2p^2(x,t_4)\bar{p}^2(x,-t_4)p_x(x,t_4)+10kp(x,t_4)p_{xx}(x,t_4)\bar{p}_x(x,-t_4)
\nonumber\\&+20k\bar{p}(x,-t_4)p_x(x,t_4)p_{xx}(x,t_4)+10kp_x^2(x,t_4)\bar{p}_x(x,-t_4),
\end{align}
where $a$ is a pure imaginary number.\\

\noindent \textbf{(3)}\, $(\varepsilon_1,\varepsilon_2)=(-1,-1)$ Reverse space-time nonlocal complex AKNS($4$) equation:
\begin{align}\label{nonlocalcomplexAKNS(4)ST}
&ap_{t_4}(x,t_4)=-p_{xxxxx}(x,t_4)+10kp\bar{p}(-x,-t_4)p_{xxx}(x,t_4)+10kp(x,t_4)p_x(x,t_4)\bar{p}_{xx}(-x,-t_4)
\nonumber\\&-30k^2p^2(x,t_4)\bar{p}^2(-x,-t_4)p_x(x,t_4)+10kp(x,t_4)p_{xx}(x,t_4)\bar{p}_x(-x,-t_4)
\nonumber\\&+20k\bar{p}(-x,-t_4)p_x(x,t_4)p_{xx}(x,t_4)+10kp_x^2(x,t_4)\bar{p}_x(-x,-t_4),
\end{align}
where $a \in \mathbb{R}$.\\

\noindent Let $\bar{g}(\varepsilon_1x,\varepsilon_2t_4)=\bar{g}^{\varepsilon}$. By Type 1 we have the Hirota bilinear forms of the equations expressed by (\ref{nonlocalcomplexAKNS(4)general}) as
\begin{align}
&(aD_{t_4}+D_x^5)\{g\cdot f\}=5kD_x\{\bar{g}^{\varepsilon}\cdot s\},\\
&D_x^2\{g\cdot g\}=fs,\\
&D_x^2\{f\cdot f\}=-2kg\bar{g}^{\varepsilon},
\end{align}
where $s$ is an auxiliary function, $a=\bar{a}\varepsilon_1\varepsilon_2$, and $f(x,t_4)=\bar{f}(\varepsilon_1x,\varepsilon_2t_4)$. \\

\section{Hirota bilinear forms of AKNS($N$) systems for $N=5, 6$}

\noindent Note that there are also some works on successive members of the AKNS hierarchy (\ref{generalsys}). But as $N$ gets larger it becomes harder to obtain the Hirota bilinear forms of AKNS($N$), $N\geq 5$. The recurrence relation (\ref{AKNS(N+1)-a})-(\ref{AKNS(N+1)-c}) is not helpful.\\

\noindent \textbf{A. AKNS($N$) system for $N=5$}\\

\noindent For $N=5$, the system (\ref{generalsys}) gives the AKNS($5$) system
\begin{align}
&ap_{t_5}=-p_{6x}+12pqp_{xxxx}+2p^2q_{xxxx}+18pq_xp_{xxx}+8pp_xq_{xxx}+30qp_xp_{xxx}+22pp_{xx}q_{xx}\nonumber\\
&-50p^2q^2p_{xx}+20qp_{xx}^2-20p^3qq_{xx}+20p_x^2q_{xx}+50p_xq_xp_{xx}-60p^2qq_xp_x-70pq^2p_x^2\nonumber\\
&-10p^3q_x^2+20p^4q^3,\label{n=5p}\\
&aq_{t_5}=q_{6x}-12pqq_{xxxx}-2q^2p_{xxxx}-18qp_xq_{xxx}-8qq_xp_{xxx}-30pq_xq_{xxx}-22qq_{xx}p_{xx}\nonumber\\
&+50q^2p^2q_{xx}-20pq_{xx}^2+20q^3pp_{xx}-20q_x^2p_{xx}-50q_xp_xq_{xx}+60q^2pp_xq_x+70qp^2q_x^2\nonumber\\
&+10q^3p_x^2-20q^4p^3. \label{n=5q}
\end{align}
Letting $p=\frac{g}{f}$ and $q=\frac{h}{f}$ in (\ref{n=5p}) and  (\ref{n=5q}) gives
\begin{align}
&(aD_{t_5}+D_x^6)\{g\cdot f\}=-10h_{xx}s+5hs_{xx}+5h_xs_x+\mu,\label{n=5-1}\\
&(aD_{t_5}-D_x^6)\{h\cdot f\}=10\tau g_{xx}-5g\tau_{xx}-5g_x\tau_x-\nu,\label{n=5-2}\\
&30hD_x\{g\cdot g_{xxx}\}+15h_xf_xs+30hf_xs_x=\mu f, \label{n=5-3}\\
&30gD_x\{h\cdot h_{xxx}\}+15\tau g_xf_x+30gf_x\tau_x=\nu f, \label{n=5-4}\\
& D_x^2\{f\cdot f \}=-2gh, \label{n=5-5}\\
& D_x^2\{g\cdot g\}=fs,\label{n=5-6}\\
&D_x^2\{h\cdot h\}=f \tau, \label{n=5-7}
\end{align}
where $s$, $\mu$, $\tau$, and $\nu$ are auxiliary functions.\\

\noindent \textbf{a. Local reductions for AKNS($5$) system}\\

\noindent Similar to the AKNS($3$) system, the AKNS($5$) system does not have $q(x,t_5)=k p(x,t_5)$ type local reduction. We consider
the local reduction $q(x,t_5)=k\bar{p}(x,t_5)$, $k$ is a real constant.\\

\noindent \textbf{a.i.} $q(x,t_5)=k\bar{p}(x,t_5)$, $k$ is a real constant.\\

\noindent The reduction $q(x,t_5)=k\bar{p}(x,t_5)$ reduces the system (\ref{n=5p}) and (\ref{n=5q}) to the local complex AKNS($5$) equation,
\begin{align}\label{localcomplexAKNS(5)}
&ap_{t_5}=-p_{6x}+12k|p|^2p_{xxxx}+2kp^2\bar{p}_{xxxx}+18kp\bar{p}_xp_{xxx}+8kpp_x\bar{p}_{xxx}+30k\bar{p}p_xp_{xxx}+22kp|p_{xx}|^2\nonumber\\
&-50k^2|p|^4p_{xx}+20k\bar{p}p_{xx}^2-20k^2p^2|p|^2\bar{p}_{xx}+20kp_x^2\bar{p}_{xx}+50k|p_x|^2p_{xx}-60k^2p|p|^2|p_x|^2\nonumber\\
&-70k^2\bar{p}|p|^2p_x^2-10k^2p^3\bar{p}_x^2+20k^3|p|^6p,
\end{align}
where $\bar{a}=-a$. This equation is known as the fifth member of reduced AKNS hierarchy or sixth order NLS equation \cite{Matveev}. Bilinear form and soliton solutions of the above equation were given in \cite{SuGao}. Multiple dromion solutions, solitons, breather, and rogue wave solutions of this equation were presented in \cite{sixthNLS1}-\cite{sixthNLS3}. If we use Type 1 approach on (\ref{n=5-1})-(\ref{n=5-7}) we obtain the Hirota bilinear form of the equation (\ref{localcomplexAKNS(5)}) as,
\begin{align}
&(aD_{t_5}+D_x^6)\{g\cdot f\}=k(-10\bar{g}_{xx}s+5\bar{g}s_{xx}+5\bar{g}_xs_x+\psi),\label{loc1n=5-1}\\
&30\bar{g}D_x\{g\cdot g_{xxx}\}+15\bar{g}_xf_xs+30\bar{g}f_xs_x=\psi f, \label{loc1n=5-2}\\
& D_x^2\{g\cdot g\}=fs,\label{loc1n=5-3}\\
& D_x^2\{f\cdot f \}=-2k|g|^2, \label{loc1n=5-4}
\end{align}
where $s$ and $\psi$ are auxiliary functions and $\bar{a}=-a$. Here $f(x,t_5)=\bar{f}(x,t_5)$.\\

\noindent \textbf{b. Nonlocal reductions for AKNS($5$) system}\\

\noindent \textbf{b.i.} $q(x,t_5)=kp(\varepsilon_1x,\varepsilon_2t_5)=kp^{\varepsilon}$, $\varepsilon_1^2=\varepsilon_2^2=1$, $k$ is a real constant.\\

\noindent This reduction reduces the system (\ref{n=5p}) and (\ref{n=5q}) to the nonlocal AKNS($5$) equations,
\begin{align} \label{nonlocalrealAKNS(5)general}
&ap_{t_5}=-p_{6x}+12kpp^{\varepsilon}p_{xxxx}+2kp^2p_{xxxx}^{\varepsilon}+18kpp_x^{\varepsilon}p_{xxx}+8kpp_xp_{xxx}^{\varepsilon}
+30kp^{\varepsilon}p_xp_{xxx}\nonumber\\
&+22kpp_{xx}p_{xx}^{\varepsilon}-50k^2p^2(p^{\varepsilon})^2p_{xx}+20kp^{\varepsilon}p_{xx}^2-20k^2p^3p^{\varepsilon}p_{xx}^{\varepsilon}
+20kp_x^2p_{xx}^{\varepsilon}
+50kp_xp_x^{\varepsilon}p_{xx}\nonumber\\
&-60k^2p^2p^{\varepsilon}p_x^{\varepsilon}p_x
-70k^2p(p^{\varepsilon})^2p_x^2-10k^2p^3(p_x^{\varepsilon})^2+20k^3p^4(p^{\varepsilon})^3,
\end{align}
where $\varepsilon_2=-1$ and $\varepsilon_1=\pm 1$. Here we have the reverse time and the reverse space-time nonlocal AKNS($5$) equations.\\

\noindent The Hirota bilinear forms of the equations (\ref{nonlocalrealAKNS(5)general}) obtained by Type 1, are given by
\begin{align}
&(aD_{t_5}+D_x^6)\{g\cdot f\}=k(-10g_{xx}^{\varepsilon}s+5g_x^{\varepsilon}s_x+5g^{\varepsilon}s_{xx}+\psi),\\
&30g^{\varepsilon}D_x\{g\cdot g_{xxx}\}+15g_x^{\varepsilon}f_xs+30g^{\varepsilon}f_xs_x=\psi f,\\
&D_x^2\{g\cdot g\}=fs,\\
& D_x^2=\{f\cdot f\}=-2gg^{\varepsilon},
\end{align}
where $s$ and $\psi$ are auxiliary functions, $g^{\varepsilon}=g(\varepsilon_1x,-t_5)$, and $f(x,t_5)=f(\varepsilon_1x,-t_5)$.\\

\noindent \textbf{b.ii.} $q(x,t_5)=k\bar{p}(\varepsilon_1x,\varepsilon_2t_5)=k\bar{p}^{\varepsilon}$, $\varepsilon_1^2=\varepsilon_2^2=1$, $k$ is a real constant.\\

\noindent By this reduction the system (\ref{n=5p}) and (\ref{n=5q}) reduces to the nonlocal complex AKNS($5$) equation,
\begin{align}\label{nonlocalcomplexAKNS(5)general}
&ap_{t_5}=-p_{6x}+12kp\bar{p}^{\varepsilon}p_{xxxx}+2kp^2\bar{p}_{xxxx}^{\varepsilon}+18kp\bar{p}_x^{\varepsilon}p_{xxx}+8kpp_x\bar{p}_{xxx}^{\varepsilon}
+30k\bar{p}^{\varepsilon}p_xp_{xxx}\nonumber\\
&+22kpp_{xx}\bar{p}_{xx}^{\varepsilon}-50k^2p^2(\bar{p}^{\varepsilon})^2p_{xx}+20k\bar{p}^{\varepsilon}p_{xx}^2-20k^2p^3\bar{p}^{\varepsilon}\bar{p}_{xx}^{\varepsilon}+20kp_x^2\bar{p}_{xx}^{\varepsilon}
+50kp_x\bar{p}_x^{\varepsilon}p_{xx}\nonumber\\
&-60k^2p^2\bar{p}^{\varepsilon}\bar{p}_x^{\varepsilon}p_x-70k^2p(\bar{p}^{\varepsilon})^2p_x^2-10k^2p^3(\bar{p}_x^{\varepsilon})^2
+20k^3p^4(\bar{p}^{\varepsilon})^3,
\end{align}
where $\bar{a}\varepsilon_2=-a$. Here we have three nonlocal complex AKNS($5$) equations; the reverse space $(\varepsilon_1,\varepsilon_2)=(-1,1)$, the reverse time $(\varepsilon_1,\varepsilon_2)=(1,-1)$, and the reverse space-time $(\varepsilon_1,\varepsilon_2)=(-1,-1)$ nonlocal equations.\\

\noindent By Type 1 that is by using $h(x,t_5)=k\bar{g}(\varepsilon_1x,\varepsilon_2t_5)=k\bar{g}^{\varepsilon}$ and $f(x,t_5)=\bar{f}(\varepsilon_1x,\varepsilon_2t_5)$ on (\ref{n=5-1})-(\ref{n=5-7}) we get Hirota bilinear forms of the equations (\ref{nonlocalcomplexAKNS(5)general}) as
\begin{align}
&(aD_{t_5}+D_x^6)\{g\cdot f\}=k(-10\bar{g}_{xx}^{\varepsilon}s+5\bar{g}_x^{\varepsilon}s_x+5\bar{g}^{\varepsilon}s_{xx}+\psi),\\
&30\bar{g}^{\varepsilon}D_x\{g\cdot g_{xxx}\}+15\bar{g}_x^{\varepsilon}f_xs+30\bar{g}^{\varepsilon}f_xs_x=\psi f,\\
&D_x^2\{g\cdot g\}=fs,\\
&D_x^2\{f\cdot f\}=-2kg\bar{g}^{\varepsilon},
\end{align}
where $s$ and $\psi$ are auxiliary functions, and $\bar{a}\varepsilon_2=-a$.\\

\noindent \textbf{B. AKNS($N$) system for $N=6$}\\

\noindent For $N=6$, the system (\ref{generalsys}) gives the AKNS($6$) system
\begin{align}
&ap_{t_6}=-p_{7x}+14pqp_{5x}+28pq_xp_{xxxx}+42qp_xp_{xxxx}+14pp_xq_{xxxx}+70qp_{xx}p_{xxx}+28pp_{xx}q_{xxx}\nonumber\\
&+98p_xq_xp_{xxx}+42pq_{xx}p_{xxx}-70p^2q^2p_{xxx}+28p_x^2q_{xxx}+112p_xp_{xx}q_{xx}-280pq^2p_xp_{xx}\nonumber\\
&-140p^2qq_xp_{xx}-140p^2qp_xq_{xx}+70p^2p_xq_x^2+140p^3q^3p_x-280pqq_xp_x^2-70q^2p_x^3,\label{n=6p}\\
&aq_{t_6}=-q_{7x}+14qpq_{5x}+28qp_xq_{xxxx}+42pq_xq_{xxxx}+14qq_xp_{xxxx}+70pq_{xx}q_{xxx}+28qq_{xx}p_{xxx}\nonumber\\
&+98q_xp_xq_{xxx}+42qp_{xx}q_{xxx}-70q^2p^2q_{xxx}+28q_x^2p_{xxx}+112q_xq_{xx}p_{xx}-280qp^2q_xq_{xx}\nonumber\\
&-140q^2pp_xq_{xx}-140q^2pq_xp_{xx}+70q^2q_xp_x^2+140q^3p^3q_x-280qpp_xq_x^2-70p^2q_x^3.\label{n=6q}
\end{align}
\noindent Letting $p=\frac{g}{f}$ and $q=\frac{h}{f}$ in (\ref{n=6p}) and  (\ref{n=6q}) gives the Hirota bilinear form of this system as
\begin{align}
&(aD_{t_6}+D_x^7)\{g\cdot f\}=14h_{xxx}s+14h_xs_{xx}-14h_{xx}s_x+\mu_1+\mu_2,\label{n=6-1}\\
&(aD_{t_6}+D_x^7)\{h\cdot f\}=14g_{xxx}\tau+14g_x\tau_{xx}-14g_{xx}\tau_{x}+\nu_1+\nu_2,\label{n=6-2}\\
&49h_xf_xs_x-28f_xh_{xx}s+35hf_xs_{xx}+98h_xD_x\{g_x\cdot g_{xx}\}+28hD_x\{g\cdot g_{xxxx}\} =-\mu_1 f, \label{n=6-3}\\
&49g_xf_x\tau_x-28f_xg_{xx}\tau+35gf_x\tau_{xx}+98g_xD_x\{h_x\cdot h_{xx}\}+28gD_x\{h\cdot h_{xxxx}\}=-\nu_1 f, \label{n=6-4}\\
&14hf_x^2s_x+14ghh_xs+7g_xh^2s+14hf_xD_x\{g\cdot g_{xxx} \}=-\mu_2f^2 \label{n=6-5}\\
&14gf_x^2\tau_x+14hgg_x\tau+7h_xg^2\tau+14gf_xD_x\{h\cdot h_{xxx} \}=-\nu_2f^2 \label{n=6-6}\\
&D_x^2\{g\cdot g\}=fs,\label{n=6-7}\\
&D_x^2\{h\cdot h\}=f \tau, \label{n=6-8}\\
&D_x^2\{f\cdot f \}=-2gh, \label{n=6-9}
\end{align}
where $s$, $\mu_1$, $\mu_2$, $\tau$, $\nu_1$, and $\nu_2$ are auxiliary functions.\\

\noindent \textbf{a. Local reductions for AKNS($6$) system}\\

\noindent \textbf{a.i.} $q(x,t_6)=kp(x,t_6)$, $k$ is a real constant.\\

\noindent When we apply the reduction $q(x,t_6)=kp(x,t_6)$ to the AKNS($6$) system (\ref{n=6p}) and (\ref{n=6q}) it reduces to the local
equation
\begin{align}\label{localrealAKNS(6)}
&ap_{t_6}=-p_{7x}+14kp^2p_{5x}+84kpp_xp_{xxxx}+140kpp_{xx}p_{xxx}-560k^2p^3p_xp_{xx}+126kp_x^2p_{xxx}\nonumber\\
&+112kp_xp_{xx}^2-70k^2p^4p_{xxx}-280k^2p^2p_x^3+140k^3p^6p_x.
\end{align}
By using Type 1 approach on (\ref{n=6-1})-(\ref{n=6-9}) we get the Hirota bilinear form of this reduced local equation as
\begin{align}
&(aD_{t_6}+D_x^7)\{g\cdot f\}=k(14g_{xxx}s+14g_xs_{xx}-14g_{xx}s_x+\psi_1+\psi_2),\label{loc1n=6-1}\\
&49g_xf_xs_x-28f_xg_{xx}s+35gf_xs_{xx}+98g_xD_x\{g_x\cdot g_{xx}\}+28gD_x\{g\cdot g_{xxxx}\}  =-\psi_1 f, \label{loc1n=6-2}\\
&14gf_x^2s_x+14kg^2g_xs+7kg_xg^2s+14gf_xD_x\{g\cdot g_{xxx} \}=-\psi_2f^2 \label{loc1n=6-3}\\
& D_x^2\{g\cdot g\}=fs,\label{loc1n=6-4}\\
& D_x^2\{f\cdot f \}=-2kg^2, \label{loc1n=6-5}
\end{align}
where $s$, $\psi_1$, and $\psi_2$ are auxiliary functions.\\

\noindent \textbf{a.ii.} $q(x,t_6)=k\bar{p}(x,t_6)$, $k$ is a real constant.\\

\noindent This system reduces to the following equation under $q(x,t_6)=k\bar{p}(x,t_6)$ consistently,
\begin{align}\label{localcomplexAKNS(6)}
&ap_{t_6}=-p_{7x}+14k|p|^2p_{5x}+28kp\bar{p}_xp_{xxxx}+42k\bar{p}p_xp_{xxxx}+14kpp_x\bar{p}_{xxxx}+70k\bar{p}p_{xx}p_{xxx}\nonumber\\
&+28kpp_{xx}\bar{p}_{xxx}+98k|p_x|^2p_{xxx}+42kp\bar{p}_{xx}p_{xxx}-70k^2|p|^4p_{xxx}+28kp_x^2\bar{p}_{xxx}+112kp_x|p_{xx}|^2\nonumber\\
&-280k^2\bar{p}|p|^2p_xp_{xx}-140k^2p|p|^2\bar{p}_xp_{xx}-140k^2p|p|^2p_x\bar{p}_{xx}+70k^2p^2\bar{p}_x|p_x|^2\nonumber\\
&+140k^3|p|^6p_x-280k^2|p|^2|p_x|^2p_x-70k^2p_x^3\bar{p}^2=0.
\end{align}
Here $a=\bar{a}$. Using Type 1 approach on (\ref{n=6-1})-(\ref{n=6-9}) gives the Hirota bilinear form of the reduced local AKNS($6$) equation as
\begin{align}
&(aD_{t_6}+D_x^7)\{g\cdot f\}=k(14\bar{g}_{xxx}s+14\bar{g}_xs_{xx}-14\bar{g}_{xx}s_x+\psi_1+\psi_2),\label{loc2n=6-1}\\
&49\bar{g}_xf_xs_x-28f_x\bar{g}_{xx}s+35\bar{g}f_xs_{xx}+98\bar{g}_xD_x\{g_x\cdot g_{xx}\}+28\bar{g}D_x\{g\cdot g_{xxxx}\}  =-\psi_1 f, \label{loc2n=6-2}\\
&14\bar{g}f_x^2s_x+14k|g|^2\bar{g}_xs+7kg_x\bar{g}^2s+14\bar{g}f_xD_x\{g\cdot g_{xxx} \}=-\psi_2f^2 \label{loc2n=6-3}\\
& D_x^2\{g\cdot g\}=fs,\label{loc2n=6-4}\\
& D_x^2\{f\cdot f \}=-2k|g|^2, \label{loc2n=6-5}
\end{align}
where $s$, $\psi_1$, and $\psi_2$ are auxiliary functions.\\

\noindent \textbf{b. Nonlocal reductions for the AKNS($6$) system}\\

\noindent \textbf{b.i.} $q(x,t_6)=kp(\varepsilon_1x,\varepsilon_2t_6)=kp^{\varepsilon}$, $\varepsilon_1^2=\varepsilon_2^2=1$, $k$ is a real constant.\\

\noindent When we apply this nonlocal reduction to the system (\ref{n=6p}) and (\ref{n=6q}), we get $\varepsilon_1=\varepsilon_2=-1$ and it reduces to the reverse space-time nonlocal AKNS($6$) equation,
\begin{align}\label{nonlocalrealAKNS(6)general}
&ap_{t_6}=-p_{7x}+14kpp^{\varepsilon}p_{5x}+28kpp_x^{\varepsilon}p_{xxxx}+42kp^{\varepsilon}p_xp_{xxxx}+14kpp_xp_{xxxx}^{\varepsilon}+70kp^{\varepsilon}p_{xx}p_{xxx}\nonumber\\
&+28kpp_{xx}p_{xxx}^{\varepsilon}+98kp_xp_x^{\varepsilon}p_{xxx}+42kpp_{xx}^{\varepsilon}p_{xxx}-70k^2p^2(p^{\varepsilon})^2p_{xxx}+28kp_x^2p_{xxx}^{\varepsilon}\nonumber\\
&+112kp_xp_{xx}p_{xx}^{\varepsilon}-280k^2p(p^{\varepsilon})^2p_xp_{xx}-140k^2p^2p^{\varepsilon}p_x^{\varepsilon}p_{xx}-140k^2p^2p^{\varepsilon}p_xp_{xx}^{\varepsilon}
\nonumber\\&+70k^2p^2p_x(p^{\varepsilon})^2+140k^3p^3(p^{\varepsilon})^3p_x-280k^2pp^{\varepsilon}p_x^{\varepsilon}p_x^2-70k^2(p^{\varepsilon})^2p_x^3,
\end{align}
where $p^{\varepsilon}=p(-x,-t_6)$.\\

\noindent By Type 1, we take $h(x,t_6)=kg(-x,-t_6)=kg^{\varepsilon}$ and $f(x,t_6)=f(-x,-t_6)$. Therefore we get the Hirota bilinear form of the equation (\ref{nonlocalrealAKNS(6)general}) as
\begin{align}
&(aD_{t_6}+D_x^7)\{g\cdot f\}=k(14g_{xxx}^{\varepsilon}s-14g_{xx}^{\varepsilon}s_x+14g_x^{\varepsilon}s_{xx}+\psi_1+\psi_2),\\
&49g_x^{\varepsilon}f_xs_x-28g_{xx}^{\varepsilon}f_xs+35g^{\varepsilon}f_xs_{xx}+98g_x^{\varepsilon}D_x\{g_x\cdot g_{xx}\}
+28g^{\varepsilon}D_x\{g\cdot g_{xxxx}\}=-\psi_1 f,\\
&14g^{\varepsilon}f_x^2s_x+14kgg^{\varepsilon}g_x^{\varepsilon}s+7kg_x(g^{\varepsilon})^2s+14g^{\varepsilon}f_xD_x\{g\cdot g_{xxx}\}=-\psi_2 f^2\\
&D_x^2\{g\cdot g\}=fs,\\
& D_x^2\{f\cdot f\}=-2gg^{\varepsilon},
\end{align}
where $s$, $\psi_1$, and $\psi_2$ are auxiliary functions.\\

\noindent \textbf{b.ii.} $q(x,t_6)=k\bar{p}(\varepsilon_1x,\varepsilon_2t_6)=k\bar{p}^{\varepsilon}$, $\varepsilon_1^2=\varepsilon_2^2=1$, $k$ is a real constant.\\

\noindent Under this reduction the system (\ref{n=6p}) and (\ref{n=6q}) reduces to the reduced nonlocal complex AKNS($6$) equation consistently,
\begin{align}\label{nonlocalcomplexAKNS(6)general}
&ap_{t_6}=-p_{7x}+14kp\bar{p}^{\varepsilon}p_{5x}+28kp\bar{p}_x^{\varepsilon}p_{xxxx}+42k\bar{p}^{\varepsilon}p_xp_{xxxx}+14kpp_x\bar{p}_{xxxx}^{\varepsilon}
+70k\bar{p}^{\varepsilon}p_{xx}p_{xxx}\nonumber\\
&+28kpp_{xx}\bar{p}_{xxx}^{\varepsilon}+98kp_x\bar{p}_x^{\varepsilon}p_{xxx}+42kp\bar{p}_{xx}^{\varepsilon}p_{xxx}-70k^2p^2(\bar{p}^{\varepsilon})^2p_{xxx}
+28kp_x^2\bar{p}_{xxx}^{\varepsilon}\nonumber\\
&+112kp_xp_{xx}\bar{p}_{xx}^{\varepsilon}-280k^2p(\bar{p}^{\varepsilon})^2p_xp_{xx}-140k^2p^2\bar{p}^{\varepsilon}\bar{p}_x^{\varepsilon}p_{xx}
-140k^2p^2\bar{p}^{\varepsilon}p_x\bar{p}_{xx}^{\varepsilon}
\nonumber\\
&+70k^2p^2p_x(\bar{p}^{\varepsilon})^2+140k^3p^3(\bar{p}^{\varepsilon})^3p_x-280k^2p\bar{p}^{\varepsilon}\bar{p}_x^{\varepsilon}p_x^2-70k^2(\bar{p}^{\varepsilon})^2p_x^3,
\end{align}
where $a=\bar{a}\varepsilon_1\varepsilon_2$. Here we have three nonlocal equations; the reverse space, the reverse time, and the reverse space-time nonlocal complex AKNS($6$) equations.\\

\noindent  By Type 1 we get the Hirota bilinear forms of the nonlocal equations expressed by (\ref{nonlocalcomplexAKNS(6)general}) as
\begin{align}
&(aD_{t_6}+D_x^7)\{g\cdot f\}=k(14\bar{g}_{xxx}^{\varepsilon}s-14\bar{g}_{xx}^{\varepsilon}s_x+14\bar{g}_x^{\varepsilon}s_x+\psi_1+\psi_2),\\
&49\bar{g}_x^{\varepsilon}f_xs_x-28\bar{g}_{xx}^{\varepsilon}f_xs+35\bar{g}^{\varepsilon}f_xs_{xx}+98\bar{g}_x^{\varepsilon}D_x\{g_x\cdot g_{xx}\}
+28\bar{g}^{\varepsilon}D_x\{g\cdot g_{xxxx}\}=-\psi_1 f\\
&14\bar{g}^{\varepsilon}f_x^2s_x+14kg\bar{g}^{\varepsilon}\bar{g}_x^{\varepsilon}s+7kg_x(\bar{g}^{\varepsilon})^2s
+14\bar{g}^{\varepsilon}f_xD_x\{g\cdot g_{xxx}\}=-\psi_2 f^2\\
&D_x^2\{g\cdot g\}=fs,\\
&D_x^2\{f\cdot f\}=-2kg\bar{g}^{\varepsilon},
\end{align}
where $s$, $\psi_1$, and $\psi_2$ are auxiliary functions, $a=\bar{a}\varepsilon_1\varepsilon_2$, $\bar{g}^{\varepsilon}=\bar{g}(\varepsilon_1x,\varepsilon_2t_6)$, and $f(x,t_6)=\bar{f}(\varepsilon_1x,\varepsilon_2t_6)$.\\

\section{Reductions, recursion operator, bilinearization, and  soliton solutions by Hirota method}

In studying the AKNS system we encounter four different operations. AKNS($N$) system has two coupled nonlinear equations of order $N+1$.
 Reduction is a process which decreases the number of dependent variables to one in a consistent way. Recursion operator increases the order of the differential equations by one. Bilinearization is a process to put the given equation, if possible, in the Hirota bilinear form. Finally, the last process is obtaining the soliton solutions of the equations by using the Hirota method. In this section we investigate how these operations commute.

\subsection{Reductions and Solutions}
We consider first the Hirota method to find the soliton solutions and reductions.
Consider the following diagram.

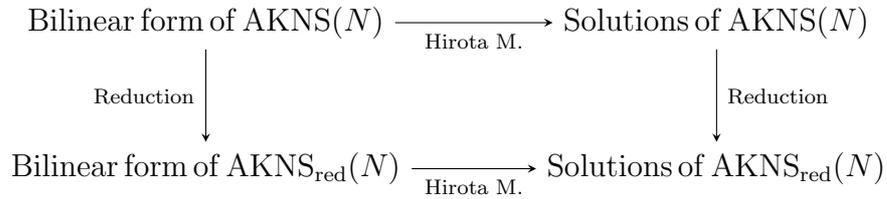
\begin{figure}[h]
\centering
\begin{tikzpicture}
  \matrix (m) [matrix of math nodes,row sep=3em,column sep=4em,minimum width=2em]
  {
     \mathrm{Bilinear\, form\, of}\, \mathrm{AKNS}(N) & \mathrm{Solutions\, of}\, \mathrm{AKNS}(N)  \\
     \mathrm{Bilinear\, form\, of}\, \mathrm{AKNS_{red}}(N) & \mathrm{Solutions\, of}\, \mathrm{AKNS_{red}}(N)  \\};
  \path[-stealth]
    (m-1-1) edge node [left] {\scriptsize{Reduction}} (m-2-1)
            edge [] node [below] {\scriptsize{Hirota M.}} (m-1-2)
    (m-2-1.east|-m-2-2) edge node [below] {\scriptsize{Hirota M.}}
            node [above] {} (m-2-2)
    (m-1-2) edge node [right] {\scriptsize{Reduction}} (m-2-2)
           ;
           \end{tikzpicture}
           \caption{Relations between reductions and solutions}
\end{figure}
\smallskip
\noindent Here $\mathrm{AKNS_{red}}(N)$ is the reduced equation obtained from the AKNS($N$) system (\ref{generalsys}) by applying a reduction.
We analyze this diagram by following two paths. Path 1 is the one starting from the bilinear form of AKNS($N$) system,
 followed by obtaining solutions of this system via Hirota method then finalized by getting solutions of the $\mathrm{AKNS_{red}}(N)$ equation by applying
 local and nonlocal reductions. Path 2 also starts from the bilinear form of the AKNS($N$) system. But the next step here is applying reduction
 to get the Hirota bilinear form of the reduced equation $\mathrm{AKNS_{red}}(N)$. The reduction is done by using Type 1 approach that has been introduced in
 \cite{GurPek1} and \cite{GurPek2}. Then we use Hirota method on this bilinear form of the reduced equation to
 obtain solutions of the $\mathrm{AKNS_{red}}(N)$ equation. Here the question is that whether we reach to the same solutions by following these two paths or not. In other words, is the above diagram commutative?\\

 Through Path 1, one can use both Type 1 and Type 2 approaches \cite{GurPek1}, \cite{GurPek3}, \cite{GurPek2} while finding soliton solutions
  of the $\mathrm{AKNS_{red}}(N)$ equation via Hirota method. These two approaches give different solutions. Clearly, through Path 2 we can only use Type 1 since we reduce the Hirota bilinear form of AKNS($N$) system to the Hirota bilinear form of $\mathrm{AKNS_{red}}(N)$ according to Type 1 approach.\\

  We obtain same solutions through both Path 1 and Path 2 if we apply Type 1 approach i.e. the diagram in Figure 2 is commutative if we use Type 1. We note that if we apply Type 1 approach, we get trivial solution for the nonlocal AKNS($2N$) or AKNS($2N+1$) equations
  reduced by $q(x,t)=kp(-x,-t)$ through each path for any $N$.

 \subsection{Reductions and Bilinearization}
 We consider now the relation between the bilinearization and the reductions.
Consider the following diagram.
\begin{figure}[h]
\centering
\begin{tikzpicture}
  \matrix (m) [matrix of math nodes,row sep=3em,column sep=4em,minimum width=2em]
  {
      \mathrm{AKNS}(N) & \mathrm{Bilinear\, form\, of}\, \mathrm{AKNS}(N)  \\
      \mathrm{AKNS_{red}}(N) & \mathrm{Bilinear\, form\, of}\, \mathrm{AKNS_{red}}(N)  \\};
  \path[-stealth]
    (m-1-1) edge node [left] {\scriptsize{Reduction}} (m-2-1)
            edge [] node [below] {\scriptsize{Transformation}} (m-1-2)
    (m-2-1.east|-m-2-2) edge node [below] {\scriptsize{Transformation}}
            node [above] {} (m-2-2)
    (m-1-2) edge node [right] {\scriptsize{Reduction}} (m-2-2)
           ;
           \end{tikzpicture}
           \caption{Relations between reductions and Hirota bilinear forms}
\end{figure}
Here we show that this diagram is commutative for $N=1, 2, 3$. Note that while applying the local and nonlocal reductions to the Hirota bilinear forms we
use Type 1.\\

\noindent \textbf{1)} \textbf{$N=1$ case:}\\

\noindent For $N=1$ we have  the AKNS($1$) system that is the coupled NLS system (\ref{n=1p}) and (\ref{n=1q}). Now we will follow two paths in the Figure 3 separately.\\

\noindent \textbf{Local reduction (a)}\, $q(x,t)=kp(x,t)$, $k$ is a real constant. There is no such local reduction in both paths. If we use it we get $a=0$.\\

\noindent \textbf{Local reduction (b)}\, $q(x,t)=k\bar{p}(x,t)$, $k$ is a real constant.\\

\noindent \noindent \textbf{Path 1.}\, If we let $p=\frac{g}{f}$ and $q=\frac{h}{f}$ in (\ref{n=1p}) and (\ref{n=1q}) we get
the Hirota bilinear form of the AKNS($1$) system as (\ref{NLShirotaa})-(\ref{NLShirotac}). Apply now the local reduction
$q(x,t)=k\bar{p}(x,t)$ to the Hirota bilinear form. Let us use Type 1 approach. This means $h=k\bar{g}$ and $f=\bar{f}$. Then we get the reduced the Hirota bilinear form as
\begin{align}
&(aD_t+D_x^2)\{g\cdot f\}=0,\label{NLSlocalredha}\\
&D_x^2\{f\cdot f\}=-2kg\bar{g},\label{NLSlocalredhb}
\end{align}
where $\bar{a}=-a$ and $f=\bar{f}$.\\

\noindent \noindent \textbf{Path 2.}\, We now first apply the reduction to the AKNS($1$) system and obtain
the reduced local equation
\begin{equation}
ap_t(x,t)=-p_{xx}(x,t)+2kp^2(x,t)\bar{p}(x,t),
 \end{equation}
 where $\bar{a}=-a$. Letting $p=\frac{g}{f}$ gives the same Hirota bilinear form of the reduced equation given in
 (\ref{NLSlocalredha}) and (\ref{NLSlocalredhb}) with $f=\bar{f}$.\\

\noindent \textbf{Nonlocal reduction (a)}\, $q(x,t)=kp(\varepsilon_1 x,\varepsilon_2t)=kp^{\varepsilon}$, $\varepsilon_1^2=\varepsilon_2^2=1$, $k$ is a real constant.\\

\noindent \textbf{Path 1.}\, We first obtain the Hirota bilinear form of the AKNS($1$) system (\ref{NLShirotaa})-(\ref{NLShirotac}) by taking $p=\frac{g}{f}$ and $q=\frac{h}{f}$ and then use the nonlocal reduction
$q(x,t)=kp(\varepsilon_1 x,\varepsilon_2t)=kp^{\varepsilon}$. This gives $h=kg^{\varepsilon}$ and $f=f^{\varepsilon}$ if we use Type 1 approach. Then the Hirota bilinear form reduces to
 \begin{align}
&(aD_t+D_x^2)\{g\cdot f\}=0,\label{NLSnonlocal1redha}\\
&D_x^2\{f\cdot f\}=-2kgg^{\varepsilon},\label{NLSnonlocal1redhb}
\end{align}
where $\varepsilon_2=-1$ and $f(x,t)=f(\varepsilon_1x,-t)$.\\

\noindent \textbf{Path 2.}\, First apply the nonlocal reduction to the AKNS($1$) system. We have the reduced nonlocal equation
\begin{equation}
ap_t(x,t)=-p_{xx}(x,t)+2kp^2(x,t)p(\varepsilon_1 x,\varepsilon_2 t),
\end{equation}
with the constraint $\varepsilon_2=-1$. Letting $p=\frac{g}{f}$ and $p^{\varepsilon}=\frac{g^{\varepsilon}}{f^{\varepsilon}}$ gives the same
Hirota bilinear form (\ref{NLSnonlocal1redha}) and (\ref{NLSnonlocal1redhb}) as in Path 1 where $f(x,t)=f(\varepsilon_1x,-t)$.\\

\noindent \textbf{Nonlocal reduction (b)}\, $q(x,t)=k\bar{p}(\varepsilon_1 x,\varepsilon_2t)=k\bar{p}^{\varepsilon}$, $\varepsilon_1^2=\varepsilon_2^2=1$, $k$ is a real constant.\\

\noindent \textbf{Path 1.}\, After obtaining the Hirota bilinear form of the AKNS($1$) system we use the nonlocal reduction
$q(x,t)=k\bar{p}(\varepsilon_1 x,\varepsilon_2t)=k\bar{p}^{\varepsilon}$ giving $h=k\bar{g}^{\varepsilon}$ and $f=\bar{f}^{\varepsilon}$ by Type 1 approach. Hence the Hirota bilinear form reduces to
 \begin{align}
&(aD_t+D_x^2)\{g\cdot f\}=0,\label{NLSnonlocal2redha}\\
&D_x^2\{f\cdot f\}=-2kg\bar{g}^{\varepsilon},\label{NLSnonlocal2redhb}
\end{align}
where $a=-\bar{a}\varepsilon_2$ and $f(x,t)=\bar{f}(\varepsilon_1x,\varepsilon_2t)$.\\

\noindent \textbf{Path 2.}\, Applying this nonlocal reduction to the AKNS($1$) system yields the reduced nonlocal equation
\begin{equation}
ap_t(x,t)=-p_{xx}(x,t)+2kp^2(x,t)\bar{p}(\varepsilon_1 x,\varepsilon_2 t),
\end{equation}
with the constraint $a=-\bar{a}\varepsilon_2$. If we take $p=\frac{g}{f}$ and $\bar{p}^{\varepsilon}=\frac{\bar{g}^{\varepsilon}}{\bar{f}^{\varepsilon}}$ we have the same
Hirota bilinear form (\ref{NLSnonlocal2redha}) and (\ref{NLSnonlocal2redhb}) as in Path 1 where $f(x,t)=\bar{f}(\varepsilon_1x,\varepsilon_2t)$.\\

\noindent \textbf{2)} \textbf{$N=2$ case:}\\

\noindent For $N=2$ we have  the AKNS($2$) system that is the coupled mKdV system (\ref{n=2p}) and (\ref{n=2q}). Now we will analyze the Figure 3.\\

\noindent \textbf{Local reduction (a)}\, $q(x,t)=kp(x,t)$, $k$ is a real constant.\\

\noindent \textbf{Path 1.}\, We first obtain the Hirota bilinear form of the AKNS($2$) system as (\ref{mKdVhirotaa})-(\ref{mKdVhirotac}) by letting
$p=\frac{g}{f}$ and $q=\frac{h}{f}$. Then we use the reduction $q(x,t)=kp(x,t)$ which gives $h=kg$. Hence the Hirota bilinear form
(\ref{mKdVhirotaa})-(\ref{mKdVhirotac}) reduces to
\begin{align}
&(aD_t+D_x^3)\{g\cdot f\}=0,\label{mKdVloc1reda}\\
&D_x^2\{f\cdot f\}=-2kg^2,\label{mKdVloc1redb}
\end{align}
without any constraint.\\

\noindent \textbf{Path 2.}\, We reduce the system to the following local reduced equation:
\begin{equation}
ap_t=-p_{xxx}+6kp^2p_x.
\end{equation}
If we let $p=\frac{g}{f}$ we get the same Hirota bilinear form (\ref{mKdVloc1reda})-(\ref{mKdVloc1redb})
for the above reduced equation.\\

\noindent \textbf{Local reduction (b)}\, $q(x,t)=k\bar{p}(x,t)$, $k$ is a real constant.\\

\noindent \textbf{Path 1.}\, After obtaining the Hirota bilinear form of the AKNS($2$) system by taking $p=\frac{g}{f}$ and $q=\frac{h}{f}$
as (\ref{mKdVhirotaa})-(\ref{mKdVhirotac}), we apply the local reduction
$q(x,t)=k\bar{p}(x,t)$. We have $h=k\bar{g}$ and $f=\bar{f}$. Then we get the reduced Hirota bilinear form as
\begin{align}
&(aD_t+D_x^3)\{g\cdot f\}=0,\label{mKdVloc2redha}\\
&D_x^2\{f\cdot f\}=-2kg\bar{g},\label{mKdVloc2redhb}
\end{align}
where $\bar{a}=a$ and $f=\bar{f}$.\\

\noindent \noindent \textbf{Path 2.}\, We  first reduce the AKNS($2$) system by using this reduction.
We get
\begin{equation}
ap_t(x,t)=-p_{xxx}(x,t)+6kp(x,t)\bar{p}(x,t)p_x,
 \end{equation}
 where $\bar{a}=a$. We then let $p=\frac{g}{f}$ and obtain the same Hirota bilinear form (\ref{mKdVloc2redha}) and (\ref{mKdVloc2redhb}) as in Path 1 for the reduced local equation  with $f=\bar{f}$.\\

\noindent \textbf{Nonlocal reduction (a)}\, $q(x,t)=kp(\varepsilon_1 x,\varepsilon_2t)=kp^{\varepsilon}$, $\varepsilon_1^2=\varepsilon_2^2=1$, $k$ is a real constant.\\

\noindent \textbf{Path 1.}\, Apply the reduction $q(x,t)=kp(\varepsilon_1 x,\varepsilon_2t)=kp^{\varepsilon}$ to the Hirota bilinear form of the AKNS($2$) system
(\ref{mKdVhirotaa})-(\ref{mKdVhirotac}) by taking $h=kg^{\varepsilon}$ and $f=f^{\varepsilon}$. The Hirota bilinear form reduces to the following bilinear form
consistently:
 \begin{align}
&(aD_t+D_x^3)\{g\cdot f\}=0,\label{mKdVnonlocal1redha}\\
&D_x^2\{f\cdot f\}=-2kgg^{\varepsilon},\label{mKdVnonlocal1redhb}
\end{align}
for $\varepsilon_1=\varepsilon_2=-1$. Here $f(x,t)=f(-x,-t)$.\\

\noindent \textbf{Path 2.}\, Reduce the AKNS($2$) system by this reduction. We only have the following reduced nonlocal equation
\begin{equation}
ap_t(x,t)=-p_{xxx}(x,t)+6kp(x,t)p(-x,-t)p_x(x,t),
\end{equation}
consistently. Now let $p=\frac{g}{f}$. We obtain the same
Hirota bilinear form (\ref{mKdVnonlocal1redha}) and (\ref{mKdVnonlocal1redhb}) as in Path 1 with $f(x,t)=f(-x,-t)$.\\

\noindent \textbf{Nonlocal reduction (b)}\, $q(x,t)=k\bar{p}(\varepsilon_1 x,\varepsilon_2t)=k\bar{p}^{\varepsilon}$, $\varepsilon_1^2=\varepsilon_2^2=1$, $k$ is a real constant.\\

\noindent \textbf{Path 1.}\, We use the reduction $q(x,t)=k\bar{p}(\varepsilon_1 x,\varepsilon_2t)=k\bar{p}^{\varepsilon}$ on the Hirota bilinear form (\ref{mKdVhirotaa})-(\ref{mKdVhirotac}) by taking $h=k\bar{g}^{\varepsilon}$ and $f=\bar{f}^{\varepsilon}$. The Hirota bilinear form reduces to
 \begin{align}
&(aD_t+D_x^3)\{g\cdot f\}=0,\label{mKdVnonlocal2redha}\\
&D_x^2\{f\cdot f\}=-2kg\bar{g}^{\varepsilon},\label{mKdVnonlocal2redhb}
\end{align}
consistently if $a=\bar{a}\varepsilon_1\varepsilon_2$. Here $f(x,t)=\bar{f}(\varepsilon_1x,\varepsilon_2t)$.\\

\noindent \textbf{Path 2.}\, Under this reduction the AKNS($2$) system reduces to
\begin{equation}
ap_t(x,t)=-p_{xxx}(x,t)+6kp(x,t)\bar{p}(\varepsilon_1 x,\varepsilon_2 t)p_x(x,t),
\end{equation}
where $a=\bar{a}\varepsilon_1\varepsilon_2$. Take $p=\frac{g}{f}$. Here again we get the same
Hirota bilinear form (\ref{mKdVnonlocal2redha}) and (\ref{mKdVnonlocal2redhb}) as in Path 1 with $f(x,t)=\bar{f}(\varepsilon_1x,\varepsilon_2t)$.\\

\noindent \textbf{3)} \textbf{$N=3$ case:}\\

\noindent For $N=3$ we have  the AKNS($3$) system (\ref{n=3p}) and (\ref{n=3q}). Let us now follow two paths in the Figure 3 separately.\\

\noindent \textbf{Local reduction (a)}\, $q(x,t)=kp(x,t)$, $k$ is a real constant. There is no such local reduction in both paths as in the AKNS($1$) case.\\

\noindent \textbf{Local reduction (b)}\, $q(x,t)=k\bar{p}(x,t)$, $k$ is a real constant.\\

\noindent \noindent \textbf{Path 1.}\, We have obtained the Hirota bilinear form of AKNS($3$) system (\ref{n=3-1})-(\ref{n=3-5}) and the reduced Hirota bilinear form (\ref{locn=3-1})-(\ref{locn=3-3}) by this reduction in Section 2. Here we get the constraint $a=-\bar{a}$ and $f(x,t)$ is a real-valued function.

\noindent \noindent \textbf{Path 2.}\, Applying this reduction to the AKNS($3$) system gives the equation (\ref{localcomplexAKNS(3)}) with $a=-\bar{a}$. If we let $p=\frac{g}{f}$, we get the same Hirota bilinear form (\ref{locn=3-1})-(\ref{locn=3-3}) as in the Path 1. Here $f=\bar{f}$.\\

\noindent \textbf{Nonlocal reduction (a)}\, $q(x,t)=kp(\varepsilon_1 x,\varepsilon_2t)=kp^{\varepsilon}$, $\varepsilon_1^2=\varepsilon_2^2=1$, $k$ is a real constant.\\

\noindent \textbf{Path 1.}\, When we apply this reduction to the Hirota bilinear form of AKNS($3$) system (\ref{n=3-1})-(\ref{n=3-5})
we get the reduced Hirota bilinear form (\ref{nonlocal1n=3-1})-(\ref{nonlocal1n=3-3}) with $a=-a\varepsilon_2$ i.e. $\varepsilon_2=-1$,
$\varepsilon_1=\pm 1$. Here $f(x,t)=f(\varepsilon_1x,-t)$.\\

\noindent \textbf{Path 2.}\, First apply the nonlocal reduction to the AKNS($3$) system. We get the reduced nonlocal equation
(\ref{nonlocalrealAKNS(3)general}) where $\varepsilon_2=-1$,
$\varepsilon_1=\pm 1$. Hence we have two types of nonlocal reduced equations here. Taking $p=\frac{g}{f}$ yields the same Hirota bilinear forms
expressed by (\ref{nonlocal1n=3-1})-(\ref{nonlocal1n=3-3}) for the
nonlocal reduced equations as in the Path 1. Here also $f(x,t)=f(\varepsilon_1x,-t)$.\\

\noindent \textbf{Nonlocal reduction (b)}\, $q(x,t)=k\bar{p}(\varepsilon_1 x,\varepsilon_2t)=k\bar{p}^{\varepsilon}$, $\varepsilon_1^2=\varepsilon_2^2=1$, $k$ is a real constant.\\

\noindent \textbf{Path 1.}\, After obtaining the Hirota bilinear form of the AKNS($3$) system which is given by (\ref{n=3-1})-(\ref{n=3-5}), we use
the nonlocal reduction
$q(x,t)=k\bar{p}(\varepsilon_1 x,\varepsilon_2t)=k\bar{p}^{\varepsilon}$ giving $h=k\bar{g}^{\varepsilon}$ and $f=\bar{f}^{\varepsilon}$. Hence the Hirota bilinear form reduces to (\ref{nonlocal2n=3-1})-(\ref{nonlocal2n=3-3}). Here $a=-\bar{a}\varepsilon_2$ and $f(x,t)=\bar{f}(\varepsilon_1x,\varepsilon_2t)$.\\

\noindent \textbf{Path 2.}\, When we apply this nonlocal reduction to the AKNS($3$) system we get three types of reduced nonlocal equations
represented by (\ref{nonlocalcomplexAKNS(3)general}) where $a=-\bar{a}\varepsilon_2$. If we take $p=\frac{g}{f}$ we have the same
Hirota bilinear form (\ref{nonlocal2n=3-1})-(\ref{nonlocal2n=3-3}) as in Path 1. Here $f(x,t)=\bar{f}(\varepsilon_1x,\varepsilon_2t)$.\\

\subsection{Reductions and Hierarchy}

The third commutativity diagram is for the reductions and the hierarchy.
\newpage
\begin{figure}[h]
\centering
\begin{tikzpicture}
  \matrix (m) [matrix of math nodes,row sep=3em,column sep=4em,minimum width=2em]
  {
     \mathrm{AKNS}(N) & \mathrm{AKNS}(N+1) \\
     \mathrm{AKNS_{red}}(N) & \mathrm{AKNS_{red}}(N+1) \\};
  \path[-stealth]
    (m-1-1) edge node [left] {\scriptsize{Reduction}} (m-2-1)
            edge [] node [below] {$\mathcal{R}$} (m-1-2)
    (m-2-1.east|-m-2-2) edge node [below] {$\mathcal{R}_{red}$}
            node [above] {} (m-2-2)
    (m-1-2) edge node [right] {\scriptsize{Reduction}} (m-2-2)
           ;
           \end{tikzpicture}
           \caption{Relation between reductions and hierarchy}
\end{figure}
Here  $\mathcal{R}$ is the recursion operator (\ref{recursion}), $\mathcal{R}_{red}$ is the same operator with a reduction applied, and
$\mathrm{AKNS_{red}}(N)$ is the AKNS($N$) system (\ref{generalsys}) with a reduction applied.

\vspace{0.3cm}

\noindent Here the question is whether the above diagram is commutative or not. In other words by starting from AKNS($N$) system and following two different paths we will check whether we reach to the same system $\mathrm{AKNS_{red}}(N+1)$ consistently or not. We begin with the case $N=1$ and check also $N=2, 3$ cases by considering all reductions -two local and two nonlocal reductions.\\

\noindent \textbf{1)} \textbf{$N=1$ case:}\\

\noindent For $N=1$ we have  the AKNS($1$) system (\ref{n=1p}) and (\ref{n=1q}). Now we will follow two paths in the Figure 4 separately.\\

\noindent \textbf{Local reduction (a)}\, $q(x,t)=kp(x,t)$, $k$ is a real constant.\\

\noindent \textbf{Path 1.}\, First, apply the recursion operator to AKNS($1$) system. We get
the AKNS($2$) system (\ref{n=2p}) and (\ref{n=2q}). Now use the local reduction $q(x,t)=kp(x,t)$ yielding
\begin{equation}\label{firstcaselocalrealredN=2}
 a \left( \begin{array}{c}
p_t  \\
kp_t
 \end{array} \right)=\left( \begin{array}{c}
-p_{xxx}+6kp^2p_x  \\
-k\frac{1}{4}p_{xxx}+6k^2p^2p_x
 \end{array} \right).
\end{equation}
 Here we have a consistent system without any additional condition.\\

\noindent \textbf{Path 2.}\, Firstly, use the local reduction $q(x,t)=kp(x,t)$ on the AKNS($1$) system (\ref{n=1p}) and (\ref{n=1q}). We have
\begin{equation}\label{firstcaselocalrealredN=1}
 a \left( \begin{array}{c}
p_t  \\
kp_t
 \end{array} \right)=\left( \begin{array}{c}
-p_{xx}+2kp^3  \\
kp_{xx}-2k^2p^3
 \end{array} \right).
\end{equation}
Notice that the above equality is satisfied if and only if $a=0$. Hence it is not possible to get the same system (\ref{firstcaselocalrealredN=2}) obtained from the Path 1.
Thus the diagram given in Figure 4 is not commutative when $N=1$ for the local reduction (a).\\

\noindent \textbf{Local reduction (b)}\, $q(x,t)=k\bar{p}(x,t)$, $k$ is a real constant.\\

\noindent \textbf{Path 1.}\, Previously, we obtain the system (\ref{n=2p}) and (\ref{n=2q}) by applying the recursion operator to AKNS($1$) system.
When we use the local reduction $q(x,t)=k\bar{p}(x,t)$ we get
\begin{equation}\label{firstcaselocalcomplexredN=2}
 a \left( \begin{array}{c}
p_t  \\
k\bar{p}_t
 \end{array} \right)=\left( \begin{array}{c}
-p_{xxx}+6kp\bar{p}p_x  \\
-k\bar{p}_{xxx}+6k^2p\bar{p}p_x
 \end{array} \right).
\end{equation}
For consistency we must have $a=\bar{a}$.\\

\noindent \textbf{Path 2.}\, We first use the local reduction $q(x,t)=k\bar{p}(x,t)$ on the AKNS($1$) system (\ref{n=1p}) and (\ref{n=1q}). We have
\begin{equation}\label{firstcaselocalcomplexredN=1}
 a \left( \begin{array}{c}
p_t  \\
k\bar{p}_t
 \end{array} \right)=\left( \begin{array}{c}
-p_{xx}+2k\bar{p}p^2  \\
k\bar{p}_{xx}-2k^2\bar{p}^2p
 \end{array} \right).
\end{equation}
Note that the above equality is satisfied if $a=-\bar{a}$ which contradicts to the condition $a=\bar{a}$ for nonzero $a$, obtained for Path 1. Therefore it is not possible to get the same system (\ref{firstcaselocalcomplexredN=2}) obtained in Path 1.
Hence the diagram given in Figure 4 is not commutative when $N=1$ for the local reduction (b).\\

\noindent \textbf{Nonlocal reduction (a)}\, $q(x,t)=kp(\varepsilon_1 x,\varepsilon_2t)=kp^{\varepsilon}$, $\varepsilon_1^2=\varepsilon_2^2=1$, $k$ is a real constant.\\

\noindent \textbf{Path 1.}\, The first step has been already done in (\ref{n=2p}) and (\ref{n=2q}). Now we use the nonlocal reduction (a) on (\ref{n=2p}) and (\ref{n=2q}), and get
\begin{equation}\label{firstcasenonlocalrealredN=2}
 a \left( \begin{array}{c}
p_t  \\
kp^{\varepsilon}_t
 \end{array} \right)=\left( \begin{array}{c}
-p_{xxx}+6kpp^{\varepsilon}p_x  \\
-kp^{\varepsilon}_{xxx}+6k^2pp^{\varepsilon}p^{\varepsilon}_x
 \end{array} \right).
\end{equation}
 This system is consistent if $a=a\varepsilon_1\varepsilon_2$ that is $\varepsilon_1=\varepsilon_2=-1$.\\

\noindent \textbf{Path 2.}\, Let us use the nonlocal reduction $q=kp^{\varepsilon}$ on the AKNS($1$) system (\ref{n=1p}) and (\ref{n=1q}). We have
\begin{equation}\label{firstcasenonlocalrealredN=1}
 a \left( \begin{array}{c}
p_t  \\
kp^{\varepsilon}_t
 \end{array} \right)=\left( \begin{array}{c}
-p_{xx}+2kp^{\varepsilon}p^2  \\
kp^{\varepsilon}_{xx}-2k^2(p^{\varepsilon})^2p
 \end{array} \right).
\end{equation}
The above system is valid if $a=-a\varepsilon_2$ that is $\varepsilon_2=-1$. Now we apply the reduced recursion operator to (\ref{firstcasenonlocalrealredN=1}). We get
\begin{align}
 a \left( \begin{array}{c}
p_t  \\
kp^{\varepsilon}_t
 \end{array} \right)&=\left( \begin{array}{cc}
2kpD^{-1}p^{\varepsilon}-D & 2pD^{-1}p  \\
-2k^2p^{\varepsilon}D^{-1}p^{\varepsilon} & -2kp^{\varepsilon}D^{-1}p+D
 \end{array} \right)\left( \begin{array}{c}
 -p_{xx}+2kp^{\varepsilon}p^2  \\
kp^{\varepsilon}_{xx}-2k^2(p^{\varepsilon})^2p
 \end{array} \right)\\
 &=\left( \begin{array}{c}
-p_{xxx}+6kpp^{\varepsilon}p_x  \\
-kp^{\varepsilon}_{xxx}+6k^2pp^{\varepsilon}p^{\varepsilon}_x
 \end{array} \right).
\end{align}
The above equality is satisfied if $a=a\varepsilon_1\varepsilon_2$ i.e. $\varepsilon_1=\varepsilon_2=-1$. Since we get the same systems in both Path 1 and Path 2 without any contradiction in constraints, we conclude that the diagram given Figure 4 is commutative when $N=1$ for the nonlocal reduction (a) with $\varepsilon_1=\varepsilon_2=-1$.\\

\noindent \textbf{Nonlocal reduction (b)}\, $q(x,t)=k\bar{p}(\varepsilon_1 x,\varepsilon_2t)=k\bar{p}^{\varepsilon}$, $\varepsilon_1^2=\varepsilon_2^2=1$, $k$ is a real constant.\\

\noindent \textbf{Path 1.}\, Use the system (\ref{n=2p}) and (\ref{n=2q}) obtained by applying recursion operator to (\ref{n=1p}) and (\ref{n=1q}). We apply the nonlocal reduction (b) on the system (\ref{n=2p}) and (\ref{n=2q}) and get
\begin{equation}\label{firstcasenonlocalcomplexredN=2}
 a \left( \begin{array}{c}
p_t  \\
k\bar{p}^{\varepsilon}_t
 \end{array} \right)=\left( \begin{array}{c}
-p_{xxx}+6kp\bar{p}^{\varepsilon}p_x  \\
-k\bar{p}^{\varepsilon}_{xxx}+6k^2p\bar{p}^{\varepsilon}\bar{p}^{\varepsilon}_x
 \end{array} \right).
\end{equation}
 This system is consistent if $a=\bar{a}\varepsilon_1\varepsilon_2$.\\

\noindent \textbf{Path 2.}\, Firstly, apply the nonlocal reduction $q=k\bar{p}^{\varepsilon}$ to the AKNS($1$) system (\ref{n=1p}) and (\ref{n=1q}). We have
\begin{equation}\label{firstcasenonlocalcomplexredN=1}
 a \left( \begin{array}{c}
p_t  \\
k\bar{p}^{\varepsilon}_t
 \end{array} \right)=\left( \begin{array}{c}
-p_{xx}+2k\bar{p}^{\varepsilon}p^2  \\
k\bar{p}^{\varepsilon}_{xx}-2k^2(\bar{p}^{\varepsilon})^2p
 \end{array} \right).
\end{equation}
The above system is consistent if $a=-\bar{a}\varepsilon_2$. Now we apply the reduced recursion operator to (\ref{firstcasenonlocalcomplexredN=1}) and obtain
\begin{align}
 a \left( \begin{array}{c}
p_t  \\
k\bar{p}^{\varepsilon}_t
 \end{array} \right)&=\left( \begin{array}{cc}
2kpD^{-1}\bar{p}^{\varepsilon}-D & 2pD^{-1}p  \\
-2k^2\bar{p}^{\varepsilon}D^{-1}\bar{p}^{\varepsilon} & -2k\bar{p}^{\varepsilon}D^{-1}p+D
 \end{array} \right)\left( \begin{array}{c}
-p_{xx}+2k\bar{p}^{\varepsilon}p^2  \\
k\bar{p}^{\varepsilon}_{xx}-2k^2(\bar{p}^{\varepsilon})^2p
 \end{array} \right)\\
 &=\left( \begin{array}{c}
-p_{xxx}+6kp\bar{p}^{\varepsilon}p_x  \\
-k\bar{p}^{\varepsilon}_{xxx}+6k^2p\bar{p}^{\varepsilon}\bar{p}^{\varepsilon}_x
 \end{array} \right).
\end{align}
The above equality is satisfied if $a=\bar{a}\varepsilon_1\varepsilon_2$. If we combine this constraint with the condition $a=-\bar{a}\varepsilon_2$ obtained previously we can conclude that the diagram given Figure 4 is commutative when $N=1$ for the nonlocal reduction (b) with $a=-\bar{a}\varepsilon_2$ and $\varepsilon_1=-1$.\\

\noindent \textbf{2)} \textbf{$N=2$ case:}\\

\noindent For $N=2$ we get the AKNS($2$) system that is coupled mKdV system (\ref{n=2p}) and (\ref{n=2q}). Now we will follow two paths in the Figure 4 separately.\\

\noindent \textbf{Local reduction (a)}\, $q(x,t)=kp(x,t)$, $k$ is a real constant.\\

\noindent \textbf{Path 1.}\, As a first step we apply the recursion operator (\ref{recursion}) to the AKNS($2$) system (\ref{n=2p}) and (\ref{n=2q}), and obtain AKNS($3$) system
\begin{equation}\label{N=3}
 a \left( \begin{array}{c}
p_t  \\
q_t
 \end{array} \right)=\left( \begin{array}{c}
-p_{xxxx}+6qp_x^2+4pp_xq_x+8pqp_{xx}+2p^2q_{xx}-6p^3q^2  \\
q_{xxxx}-6pq_x^2-4qp_xq_x-8pqq_{xx}-2q^2p_{xx}+6p^2q^3
 \end{array} \right).
 \end{equation}
Now apply the local reduction (a) to the above system. We get
\begin{equation}\label{secondcaselocalrealredN=3}
 a \left( \begin{array}{c}
p_t  \\
kp_t
 \end{array} \right)=\left( \begin{array}{c}
-p_{xxxx}+10kpp_x^2+10kp^2p_{xx}-6k^2p^5 \\
kp_{xxxx}-10k^2pp_x^2-10k^2p^2p_{xx}+6k^3p^5
 \end{array} \right).
 \end{equation}
This equality is valid only if $a=0$. Therefore there is not such a consistent local reduction.\\

\noindent \textbf{Path 2.}\, Use first the reduction on the AKNS($2$) system (\ref{n=2p}) and (\ref{n=2q}). We have
\begin{equation}\label{secondcaselocalrealredN=2}
 a \left( \begin{array}{c}
p_t  \\
kp_t
 \end{array} \right)=\left( \begin{array}{c}
-p_{xxx}+6kp^2p_x  \\
-kp_{xxx}+6k^2p^2p_x
 \end{array} \right).
\end{equation}
This system is consistent without any condition. As a second step, we apply the reduced recursion operator to the above system and get
again the system (\ref{secondcaselocalrealredN=3}) which is valid only if $a=0$. Hence in either path we do not have a consistent local reduction (a).\\

\noindent \textbf{Local reduction (b)}\, $q(x,t)=k\bar{p}(x,t)$, $k$ is a real constant.\\

\noindent \textbf{Path 1.}\, Previously we have obtained AKNS($3$) system (\ref{N=3}).
Now we apply the local reduction (b) to that system and get
\begin{equation}\label{secondcaselocalcomplexredN=3}
 a \left( \begin{array}{c}
p_t  \\
k\bar{p}_t
 \end{array} \right)=\left( \begin{array}{c}
-p_{xxxx}+6k\bar{p}p_x^2+4kpp_x\bar{p}_x+8kp\bar{p}p_{xx}+2kp^2\bar{p}_{xx}-6k^2p^3\bar{p}^2 \\
k\bar{p}_{xxxx}-6k^2p\bar{p}_x^2-4k^2\bar{p}p_x\bar{p}_x-8k^2p\bar{p}\bar{p}_{xx}
-2k^2\bar{p}^2p_{xx}+2k^3p^2\bar{p}^3
 \end{array} \right).
 \end{equation}
This system is consistent if $a=-\bar{a}$.\\

\noindent \textbf{Path 2.}\, We use first the reduction on (\ref{n=2p}) and (\ref{n=2q}), and we have
\begin{equation}\label{secondcaselocalcomplexredN=2}
 a \left( \begin{array}{c}
p_t  \\
k\bar{p}_t
 \end{array} \right)=\left( \begin{array}{c}
-p_{xxx}+6kp\bar{p}p_x  \\
-k\bar{p}_{xxx}+6k^2p\bar{p}\bar{p}_x
 \end{array} \right),
\end{equation}
which is valid if $a=\bar{a}$. If we also apply the reduced recursion operator to that system and we again obtain
the system (\ref{secondcaselocalcomplexredN=3}) which is valid only if $a=-\bar{a}$. These two conditions on $a$ yields $a=0$. Therefore the results obtained from Path 1 and Path 2 do not coincide. Hence the diagram given in Figure 4 is not commutative when $N=2$ for the local reduction (b).\\

\noindent \textbf{Nonlocal reduction (a)}\, $q(x,t)=kp(\varepsilon_1 x,\varepsilon_2t)=kp^{\varepsilon}$, $\varepsilon_1^2=\varepsilon_2^2=1$, $k$ is a real constant.\\

\noindent \textbf{Path 1.}\, After obtaining the AKNS($3$) system (\ref{N=3}), we apply the nonlocal reduction (a) to that system. We get
\begin{equation}\label{secondcasenonlocalrealredN=3}
 a \left( \begin{array}{c}
p_t  \\
kp^{\varepsilon}_t
 \end{array} \right)=\left( \begin{array}{c}
p_{xxxx}+6kp^{\varepsilon}p_x^2+4kpp_xp^{\varepsilon}_x+8kpp^{\varepsilon}p_{xx}
+2kp^2p^{\varepsilon}_{xx}-6k^2p^3(p^{\varepsilon})^2 \\
kp^{\varepsilon}_{xxxx}-6k^2p(p^{\varepsilon}_x)^2-4k^2p^{\varepsilon}p^{\varepsilon}_xp_x-8k^2p^{\varepsilon}pp^{\varepsilon}_{xx}
-2k^2(p^{\varepsilon})^2p_{xx}+6k^3(p^{\varepsilon})^3p^2
 \end{array} \right).
 \end{equation}
This equality is valid if $a=-a\varepsilon_2$ that is $\varepsilon_2=-1$.

\noindent \textbf{Path 2.}\, At first use the nonlocal reduction (a) on AKNS($2$) system (\ref{n=2p}) and (\ref{n=2q}), and get
\begin{equation}\label{secondcasenonlocalrealredN=2}
 a \left( \begin{array}{c}
p_t  \\
kp^{\varepsilon}_t
 \end{array} \right)=\left( \begin{array}{c}
-p_{xxx}+6kpp^{\varepsilon}p_x  \\
-kp^{\varepsilon}_{xxx}+6k^2pp^{\varepsilon}p^{\varepsilon}_x
 \end{array} \right).
\end{equation}
The above system is consistent if $a=a\varepsilon_1\varepsilon_2$ yielding $\varepsilon_1=\varepsilon_2=-1$. In addition to that if we also apply the
reduced recursion operator to this system we get (\ref{secondcasenonlocalrealredN=3}) which is valid if $a=-a\varepsilon_2$ i.e. $\varepsilon_2=-1$. Since we get the same systems in both Path 1 and Path 2 without any contradiction in constraints, we conclude that the diagram given Figure 4 is commutative when $N=2$ for the nonlocal reduction (a) with $\varepsilon_1=\varepsilon_2=-1$.\\

\noindent \textbf{Nonlocal reduction (b)}\, $q(x,t)=k\bar{p}(\varepsilon_1 x,\varepsilon_2t)=k\bar{p}^{\varepsilon}$, $\varepsilon_1^2=\varepsilon_2^2=1$, $k$ is a real constant.\\

\noindent \textbf{Path 1.}\,  After obtaining the AKNS($3$) system (\ref{N=3}), we apply the nonlocal reduction (b) to that system. We have
\begin{equation}\label{secondcasenonlocalcomplexredN=3}
 a \left( \begin{array}{c}
p_t  \\
k\bar{p}^{\varepsilon}_t
 \end{array} \right)=\left( \begin{array}{c}
p_{xxxx}+6k\bar{p}^{\varepsilon}p_x^2+4kpp_x\bar{p}^{\varepsilon}_x+8kp\bar{p}^{\varepsilon}p_{xx}
+2kp^2\bar{p}^{\varepsilon}_{xx}-6k^2p^3(\bar{p}^{\varepsilon})^2 \\
k\bar{p}^{\varepsilon}_{xxxx}-6k^2p(\bar{p}^{\varepsilon}_x)^2-4k^2\bar{p}^{\varepsilon}\bar{p}^{\varepsilon}_xp_x
-8k^2\bar{p}^{\varepsilon}p\bar{p}^{\varepsilon}_{xx}
-2k^2(\bar{p}^{\varepsilon})^2p_{xx}+6k^3(\bar{p}^{\varepsilon})^3p^2
 \end{array} \right).
 \end{equation}
This system is consistent if $a=-\bar{a}\varepsilon_2$.\\

\noindent \textbf{Path 2.}\, We first use the nonlocal reduction (b) on AKNS($2$) system (\ref{n=2p}) and (\ref{n=2q}), and obtain
\begin{equation}\label{secondcasenonlocalcomplexredN=2}
 a \left( \begin{array}{c}
p_t  \\
k\bar{p}^{\varepsilon}_t
 \end{array} \right)=\left( \begin{array}{c}
-p_{xxx}+6kp\bar{p}^{\varepsilon}p_x  \\
-k\bar{p}^{\varepsilon}_{xxx}+6k^2p\bar{p}^{\varepsilon}\bar{p}^{\varepsilon}_x
 \end{array} \right).
\end{equation}
The above system is consistent if $a=\bar{a}\varepsilon_1\varepsilon_2$. After that we apply the
reduced recursion operator to this system and we get (\ref{secondcasenonlocalcomplexredN=3}) which is valid if $a=-\bar{a}\varepsilon_2$. If we combine this constraint with the condition $a=\bar{a}\varepsilon_1\varepsilon_2$ obtained previously we can conclude that the diagram given Figure 4 is commutative when $N=1$ for the nonlocal reduction (b) with $a=-\bar{a}\varepsilon_2$ and $\varepsilon_1=-1$.\\

\noindent \textbf{3)} \textbf{$N=3$ case:}\\

\noindent For $N=3$ we have the AKNS($3$) system (\ref{n=3p}) and (\ref{n=3q}). Now we will follow two paths in the Figure 1 separately.\\

\noindent \textbf{Local reduction (a)}\, $q(x,t)=kp(x,t)$, $k$ is a real constant.\\

\noindent \textbf{Path 1.}\, We apply the recursion operator (\ref{recursion}) to the AKNS($3$) system (\ref{n=3p}) and (\ref{n=3q}) and obtain AKNS($4$) system
(\ref{n=4p}) and (\ref{n=4q}). When we use the local reduction (a) on the above system we get
\begin{equation}\label{thirdcaselocalrealredN=4}
 a \left( \begin{array}{c}
p_t  \\
kp_t
 \end{array} \right)=\left( \begin{array}{c}
-p_{xxxxx}+10kp^2p_{xxx}+40kpp_xp_{xx}-30k^2p^4p_x+10kp_x^3 \\
-kp_{xxxxx}+10k^2p^2p_{xxx}+40k^2pp_xp_{xx}-30k^3p^4p_x+10k^2p_x^3
 \end{array} \right).
 \end{equation}
This system is consistent without any condition.\\

\noindent \textbf{Path 2.}\, We apply first the reduction on the AKNS($3$) system (\ref{n=3p}) and (\ref{n=3q}), and have
\begin{equation}\label{thirdcaselocalrealredN=3}
 a \left( \begin{array}{c}
p_t  \\
kp_t
 \end{array} \right)=\left( \begin{array}{c}
-p_{xxxx}+10kpp_x^2+10kp^2p_{xx}-6k^2p^5  \\
kp_{xxxx}-10k^2pp_x^2-10k^2p^2p_{xx}+6k^3p^5
 \end{array} \right).
\end{equation}
Obviously, the above equality is satisfied only when $a=0$. However if we apply the reduced recursion operator to the above system we
get (\ref{thirdcaselocalrealredN=4}) as in Path 1. But since the first step of Path 2 gives a system which is not consistent we can conclude that
the diagram given in Figure 4 is not commutative when $N=3$ for the local reduction (a) which is similar to the case when $N=1$.\\

\noindent \textbf{Local reduction (b)}\, $q(x,t)=k\bar{p}(x,t)$, $k$ is a real constant.\\

\noindent \textbf{Path 1.}\, Use the AKNS($4$) system (\ref{n=4p}) and (\ref{n=4q}), and apply the local reduction (b) to that system. We have
\begin{equation}\label{thirdcaselocalcomplexredN=4}
 a \left( \begin{array}{c}
p_t  \\\\
k\bar{p}_t
 \end{array} \right)=\left( \begin{array}{c}
-p_{xxxxx}+10kp\bar{p}p_{xxx}+10kpp_x\bar{p}_{xx}-30k^2p^2\bar{p}^2p_x+10kpp_{xx}\bar{p}_x
\\\hfill+20k\bar{p}p_xp_{xx}+10kp_x^2\bar{p}_x \\
-k\bar{p}_{xxxxx}+10k^2p\bar{p}p_{xxx}+10k^2pp_x\bar{p}_{xx}-30k^3p^2\bar{p}^2p_x+10k^2pp_{xx}\bar{p}_x
\\\hfill+20k^2\bar{p}p_xp_{xx}+10k^2p_x^2\bar{p}_x
 \end{array} \right).
 \end{equation}
This system is consistent if $a=\bar{a}$.\\

\noindent \textbf{Path 2.}\, At first use the reduction on (\ref{n=3p}) and (\ref{n=3q}). We have
\begin{equation}\label{thirdcaselocalcomplexredN=3}
 a \left( \begin{array}{c}
p_t  \\
k\bar{p}_t
 \end{array} \right)=\left( \begin{array}{c}
-p_{xxxx}+6k\bar{p}p_x^2+4kpp_x\bar{p}_x+8kp\bar{p}p_{xx}+2kp^2\bar{p}_{xx}-6k^2p^3\bar{p}^2 \\
k\bar{p}_{xxxx}-6k^2p\bar{p}_x^2-4k^2\bar{p}\bar{p}_xp_x-8k^2\bar{p}p\bar{p}_{xx}-2k^2\bar{p}^2p_{xx}+6k^3\bar{p}^3p^2
 \end{array} \right),
\end{equation}
which is valid if $a=-\bar{a}$. Then we apply the reduced recursion operator to that system and we again obtain
the system (\ref{thirdcaselocalcomplexredN=4}) which is consistent if $a=\bar{a}$. These two conditions on the constant $a$ yields $a=0$. Therefore the results obtained from Path 1 and Path 2 do not coincide. Hence the diagram given in Figure 4 is not commutative when $N=3$ for the local reduction (b) as in the case $N=1$.\\

\noindent \textbf{Nonlocal reduction (a)}\, $q(x,t)=kp(\varepsilon_1 x,\varepsilon_2t)=kp^{\varepsilon}$, $\varepsilon_1^2=\varepsilon_2^2=1$, $k$ is a real constant.\\

\noindent \textbf{Path 1.}\, After obtaining the AKNS($4$) system (\ref{n=4p}) and (\ref{n=4q}), applying the nonlocal reduction (a) to that system gives
\begin{equation}\label{thirdcasenonlocalrealredN=4}
 a \left( \begin{array}{c}
p_t  \\\\
kp^{\varepsilon}_t
 \end{array} \right)=\left( \begin{array}{c}
-p_{xxxxx}+10kpp^{\varepsilon}p_{xxx}+10kpp_xp^{\varepsilon}_{xx}-30k^2p^2(p^{\varepsilon})^2p_x+10kpp_{xx}p^{\varepsilon}_x
\\\hfill+20kp^{\varepsilon}p_xp_{xx}+10kp_x^2p^{\varepsilon}_x \\
-kp^{\varepsilon}_{xxxxx}+10k^2pp^{\varepsilon}p^{\varepsilon}_{xxx}+10k^2p^{\varepsilon}p^{\varepsilon}_xp_{xx}
-30k^3p^2(p^{\varepsilon})^2p^{\varepsilon}_x+10k^2p^{\varepsilon}p^{\varepsilon}_{xx}p_x\\\hfill+20k^2pp^{\varepsilon}_xp^{\varepsilon}_{xx}
+10k^2(p^{\varepsilon}_x)^2p_x
 \end{array} \right).
 \end{equation}
This system is consistent if $a=a\varepsilon_1\varepsilon_2$ i.e. $\varepsilon_1=\varepsilon_2=-1$.\\

\noindent \textbf{Path 2.}\, At first use the nonlocal reduction (a) on AKNS($3$) system (\ref{n=3p}) and (\ref{n=3q}), and get
\begin{equation}\label{thirdcasenonlocalrealredN=3}
 a \left( \begin{array}{c}
p_t  \\
kp^{\varepsilon}_t
 \end{array} \right)=\left( \begin{array}{c}
-p_{xxxx}+6kp^{\varepsilon}p_x^2+4kpp_xp^{\varepsilon}_x+8kpp^{\varepsilon}p_{xx}+2kp^2p^{\varepsilon}_{xx}-6k^2p^3(p^{\varepsilon})^2  \\
+kp^{\varepsilon}_{xxxx}-6k^2p(p^{\varepsilon}_x)^2-4k^2p^{\varepsilon}p^{\varepsilon}_xp_x-8k^2pp^{\varepsilon}p^{\varepsilon}_{xx}
-2k^2(p^{\varepsilon})^2p_{xx}+6k^3(p^{\varepsilon})^3p^2
 \end{array} \right).
\end{equation}
This system is consistent if $a=-a\varepsilon_2$ yielding $\varepsilon_2=-1$. When we apply the reduced recursion operator to the above system we get
the same system (\ref{thirdcasenonlocalrealredN=4}) which is valid if $\varepsilon_1=\varepsilon_2=-1$. Hence we conclude that the diagram given Figure 4 is commutative when $N=3$ for the nonlocal reduction (a) with $\varepsilon_1=\varepsilon_2=-1$.\\

\noindent \textbf{Nonlocal reduction (b)}\, $q(x,t)=k\bar{p}(\varepsilon_1 x,\varepsilon_2t)=k\bar{p}^{\varepsilon}$, $\varepsilon_1^2=\varepsilon_2^2=1$, $k$ is a real constant.\\

\noindent \textbf{Path 1.}\,  We apply the nonlocal reduction (b) to the AKNS($4$) system (\ref{n=4p}) and (\ref{n=4q}). We have
\begin{equation}\label{thirdcasenonlocalcomplexredN=4}
 a \left( \begin{array}{c}
p_t  \\
k\bar{p}^{\varepsilon}_t
 \end{array} \right)=\left( \begin{array}{c}
-p_{xxxxx}+10kp\bar{p}^{\varepsilon}p_{xxx}+10kpp_x\bar{p}^{\varepsilon}_{xx}-30k^2p^2(\bar{p}^{\varepsilon})^2p_x
+10kpp_{xx}\bar{p}^{\varepsilon}_x
\\\hfill+20k\bar{p}^{\varepsilon}p_xp_{xx}+10kp_x^2\bar{p}^{\varepsilon}_x \\
-k\bar{p}^{\varepsilon}_{xxxxx}+10k^2p\bar{p}^{\varepsilon}\bar{p}^{\varepsilon}_{xxx}+10k^2\bar{p}^{\varepsilon}\bar{p}^{\varepsilon}_xp_{xx}
-30k^3p^2(\bar{p}^{\varepsilon})^2\bar{p}^{\varepsilon}_x+10k^2\bar{p}^{\varepsilon}\bar{p}^{\varepsilon}_{xx}p_x\\\hfill
+20k^2p\bar{p}^{\varepsilon}_x\bar{p}^{\varepsilon}_{xx}
+10k^2(\bar{p}^{\varepsilon}_x)^2p_x
 \end{array} \right).
 \end{equation}
This system is consistent if $a=\bar{a}\varepsilon_1\varepsilon_2$.\\

\noindent \textbf{Path 2.}\, Firstly, we use the nonlocal reduction (b) on AKNS($3$) system (\ref{n=3p}) and (\ref{n=3q}), and get
\begin{equation}\label{thirdcasenonlocalcomplexredN=3}
 a \left( \begin{array}{c}
p_t  \\
k\bar{p}^{\varepsilon}_t
 \end{array} \right)=\left( \begin{array}{c}
-p_{xxxx}+6k\bar{p}^{\varepsilon}p_x^2+4kpp_x\bar{p}^{\varepsilon}_x+8kp\bar{p}^{\varepsilon}p_{xx}
+2kp^2\bar{p}^{\varepsilon}_{xx}-6k^2p^3(\bar{p}^{\varepsilon})^2  \\
k\bar{p}^{\varepsilon}_{xxxx}
-6k^2p(\bar{p}^{\varepsilon}_x)^2-4k^2\bar{p}^{\varepsilon}\bar{p}^{\varepsilon}_xp_x-8k^2p\bar{p}^{\varepsilon}\bar{p}^{\varepsilon}_{xx}
-2k^2(\bar{p}^{\varepsilon})^2p_{xx}+6k^3(\bar{p}^{\varepsilon})^3p^2
 \end{array} \right).
\end{equation}
The above system is consistent if $a=-\bar{a}\varepsilon_2$. When we apply the reduced recursion operator to that system we get
again the system (\ref{thirdcasenonlocalcomplexredN=4}) which is consistent if $a=\bar{a}\varepsilon_1\varepsilon_2$. If we combine both of these constraints and consider also Path 1,
we conclude that the diagram given by Figure 4 is commutative when $N=3$ for the nonlocal reduction (b) with $a=-\bar{a}\varepsilon_2$ and $\varepsilon_1=-1$.\\

\section{Conclusion}
In this work we considered the AKNS($N$) hierarchy for $N=3,4,5,6$. We gave the Hirota bilinear forms of these systems. The Hirota bilinear forms of these systems are different then the ones for AKNS($1$) and AKNS($2$) systems. They are indeed inhomogeneous and hard to obtain from the recurrence relation given
previously. We presented the local and nonlocal reductions of the AKNS($3$), AKNS($4$), AKNS($5$), and AKNS($6$) systems. We gave also the Hirota bilinear forms of the reduced local and nonlocal equations.

In studying Hirota bilinearization and reduction of the AKNS($N$) systems we have analyzed commutativity diagrams of the operations involved in. These are recursion operator, reduction of the systems, and Hirota bilinearization. All these diagrams turn out to be compatible under certain conditions.

\section{Acknowledgment}
  This work is partially supported by the Scientific
and Technological Research Council of Turkey (T\"{U}B\.{I}TAK).\\

\end{document}